%% file: ms.tex
\documentclass[12pt]{article}

\usepackage{amsmath}
\usepackage{scicite}
\usepackage{times}
\usepackage{graphicx}
\usepackage{longtable}
\usepackage{siunitx}
\usepackage{amssymb}
\usepackage{amsmath}
\usepackage{verbatim}
\usepackage{ccaption}
\usepackage[colorinlistoftodos]{todonotes}

\newcommand{\degr}{^{\circ}}

\newcommand{\tmop}[1]{\ensuremath{\operatorname{#1}}}
\newcommand{\beginsupplement}{%
        \setcounter{table}{0}
        \renewcommand{\thetable}{S\arabic{table}}%
        \setcounter{figure}{0}
        \renewcommand{\thefigure}{S\arabic{figure}}%
}

\newcommand\aap{Astron.~Astrophys.}                % Astronomy and Astrophysics  
\newcommand\apjl{Astrophys.~J.}                % Astrophysical Journal, Letters
\newcommand\apjs{Astrophys.~J.~Suppl.~Ser.}               % Astrophysical Journal, Supplement
\newenvironment{sciabstract}{
\begin{quote} \bf}
{\end{quote}}
\newcounter{lastnote}

\title{Radio emission from a pulsar's magnetic pole revealed by general relativity}
\author
{Gregory Desvignes,$^{1,2\ast}$ Michael Kramer,$^{1,3}$ Kejia Lee,$^{4}$ Joeri van Leeuwen,$^{5,6}$\\
 Ingrid~Stairs,$^{7}$
 Axel~Jessner,$^{1}$
 Isma\"el~Cognard,$^{8,9}$
 Laura~Kasian,$^{7}$\\
 Andrew~Lyne,$^3$
 Ben~W.~Stappers$^3$\\
\\
\normalsize{$^{1}$Max-Planck-Institut f\"ur Radioastronomie, Auf dem H\"ugel, 69 D-53121 Bonn, Germany}\\
\normalsize{$^{2}$Laboratoire d'\'Etudes Spatiales et d'Instrumentation en Astrophysique, Observatoire de Paris,}\\
\normalsize{Universit\'e Paris-Sciences-et-Lettres, Centre National de la Recherche Scientifique, Sorbonne Universit\'e,}\\
\normalsize{Universit\'e de Paris, 5 place Jules Janssen, 92195 Meudon, France}\\
\normalsize{$^{3}$Jodrell Bank Centre for Astrophysics, School of Physics and Astronomy,}\\
\normalsize{The University of Manchester, Manchester M13 9PL, UK}\\
\normalsize{$^{4}$Kavli institute for astronomy and astrophysics, Peking University,}\\
\normalsize{Beijing 100871, People's Republic of China}\\
\normalsize{$^{5}$ASTRON, The Netherlands Institute for Radio Astronomy,}\\
\normalsize{Postbus 2, 7990 AA Dwingeloo, The Netherlands}\\
\normalsize{$^{6}$Astronomical Institute Anton Pannekoek, University of Amsterdam,}\\
\normalsize{Science Park 904, 1098 XH Amsterdam, The Netherlands}\\
\normalsize{$^{7}$Department of Physics and Astronomy, University of British Columbia,}\\
\normalsize{Vancouver, BC V6T 1Z1, Canada}\\
\normalsize{$^{8}$Laboratoire de Physique et Chimie de l'Environnement et de l'Espace,}\\
\normalsize{Centre National de la Recherche Scientifique-Universit{\'e} d'Orl{\'e}ans, F-45071 Orl{\'e}ans, France}\\
\normalsize{$^{9}$Station de radioastronomie de Nan{\c c}ay, Observatoire de Paris,}\\
\normalsize{Centre National de la Recherche Scientifique,}\\
\normalsize{Institut national des sciences de l'Univers, F-18330 Nan{\c c}ay, France}\\
\\
\normalsize{$^\ast$To whom correspondence should be addressed; E-mail:  gdesvignes@mpifr-bonn.mpg.de.}
}

\date{}
\begin{document}
\baselineskip24pt

\maketitle

% Abstract should be < 125 words
% All figure legends should be < 200 words
\begin{sciabstract}
% Observations of relativistic spin-precession in binary pulsars allow
%for a wide range of studies from tests of General Relativity (GR) to
%constraining the pulsar emission mechanism. 
Binary pulsars are affected by general relativity (GR), causing the
spin axes of each pulsar to precess. We present polarimetric radio
observations of PSR J1906$+$0746 that demonstrate the validity of the
geometrical model of pulsar polarisation.  We reconstruct the
(sky-projected) polarisation emission map over the pulsar's magnetic
pole and predict the disappearance of the detectable emission by
2028. Two additional tests of GR are performed in this system,
including the spin-precession for strongly self-gravitating bodies. We
constrain the relativistic treatment of the pulsar polarisation model
and measure the pulsar beaming fraction, with implications for the
population of neutron stars and the expected rate of neutron star
mergers.
\end{sciabstract}

%\section*{Introduction}
Pulsars are fast-spinning neutron stars of mass $\sim 1.2-2.2$ solar
masses (M$_{\odot}$) with strong magnetic fields that emit a beam of
radio waves along their magnetic axes above each of their opposite
magnetic poles.  Einstein's theory of general relativity (GR) predicts
that space-time is curved by massive bodies. Predicted effects of this
include relativistic spin-precession in binary pulsars
\cite{dr74}.  This precession arises from any misalignment, by an
angle $\delta$, of the spin vector of each pulsar with respect to the
total angular momentum vector of the binary, most likely caused by an
asymmetric supernova explosion imparting a kick onto (one of) the
neutron star(s) \cite{wkk00,jan17}. This precession causes the viewing
geometry to vary, which can be tested observationally.
% Studies can provide
%information on neutron star formation, population studies, neutron
%star mergers and gravitational wave detection rates, tests of theories
%of gravity, or insight into the little-understood radio pulsar
%emission itself.
%We have observed PSR J1906$+$0746, a young binary pulsar with a 4-hr
%orbit; radio emission is detected from one of the pulsar with spin
%period $P_\textrm{S} \sim 144$\,ms.
% in a 4-hr orbit around an
%other neutron star \cite{lsf+06,vks+15}, provides the opportunity to
%scrutinize all of these aspects in a single source in a
%self-consistent way.

As a pulsar rotates, its radio beams sweep the sky. If one of the
beams crosses our line of sight (LOS), its emission is perceived as
being pulsed, that, when averaged over several hundreds of pulsar
rotations, typically forms a stable pulse profile. Evidence for a
variable pulse profile attributed to changes in the viewing geometry
caused by spin-precession have been observed and modelled for the
binary pulsar B1913$+$16 \cite{wrt89,kra98}.  Polarisation information
can provide an additional and independent tool to study relativistic
spin-precession \cite{dt92,sta04}. We expect that the position angle
(PA) sweep of a pulsar's linearly polarised emission due to
geometrical effects can be described by the Rotating Vector Model
(RVM) \cite{rc69}.  This simple model, which assumes a magnetic dipole
centred on the pulsar, relates the PA to the projection of the
magnetic field line direction as the pulsar beam rotates and crosses
our LOS. The resulting gradient of the PA sweep as a function of the
pulsar rotational phase \cite{k70} depends only on the magnetic
inclination angle $\alpha$ and the impact parameter $\beta$ (i.e. the
angle of closest approach between the observer direction and the
magnetic axis). For LOSs crossing opposite sides of the same magnetic
pole, the RVM predicts opposite slopes of the PA swing, which becomes
steeper for smaller $\beta$ \cite{lk05}.  RVM has been extended to
include rotational and relativistic effects between the pulsar and
observer frame \cite{bcw91,dyk08}, in principle allowing emission
heights for the observed radio emission to be estimated. Although the
RVM matches observations of young pulsars that present a smooth PA
swing \cite{kj08}, deviations are also observed, so for large emission
heights the pure dipole approximation may not be valid for a rotating
plasma-loaded magnetosphere \cite{ae98}. Evidence for the central
assumption of all these models, i.e. the geometrical meaning of the PA
sweep, has so far been missing.

%We have observed PSR J1906$+$0746 (position
%$19^h06^m48.86^s+07^\circ46'25.9''$, J2000 equinox), a young binary
%pulsar with a 4-hr orbit; radio emission is detected from one of the
%qpulsar with spin period $P_\textrm{S} \sim 144$\,ms. 

When PSR J1906$+$0746 (position $19^h06^m48.86^s+07^\circ46'25.9''$,
J2000 equinox), a young pulsar with spin period $P_\textrm{S} \sim
144$\,ms in a 4-hr orbit around an other neutron star, was discovered
in 2004 \cite{lsf+06}, it showed two polarised emission components
separated by nearly half a period (or $\sim 180$ deg of pulse
longitude). The ``main pulse'' (MP) and ``interpulse'' (IP) indicated
a nearly orthogonal geometry where emission from both magnetic poles
is visible from Earth. Comparison with archival data from the Parkes
Multibeam Pulsar Survey (PMPS) \cite{mlc+01} revealed that only the
stronger MP had been visible in 1998 \cite{lsf+06}, suggesting effects
of relativistic spin precession.  Timing analysis of this pulsar
\cite{vks+15} has measured three relativistic corrections to the
orbit, the Post-Keplerian (PK) parameters. Two of the PK parameters,
the periastron advance $\dot\omega$ and the time dilation $\gamma$,
allow us to determine a pulsar and companion mass of
$m_{\textrm{p}}=1.29\pm0.01$~M$_{\odot}$ and
$m_{\textrm{c}}=1.32\pm0.01$~M$_{\odot}$, respectively, assuming GR is
correct. The third PK parameter, the orbital decay due to gravitational
wave emission $\dot{P}_{\textrm b}$, is consistent with these
measurements, providing a 5\% test of GR \cite{vks+15}. GR also
provides an estimate of the orbital inclination angle
$i=43.7\pm0.4\degr$ or $136.3\pm0.4\degr$ (due to a $180^\circ$
degeneracy) and the spin precession rate
$\Omega_{\textrm{p}}=2.234\pm0.014\degr$\,yr$^{-1}$.  Independent
measurements of these two terms are potentially observable in
precessing systems, allowing these GR predicted values to be
tested. Observations of the Double Pulsar \cite{bkk+08} and
PSR~B1534+12 \cite{sta04,fst14} match the GR predictions of
spin-precession with precisions of 13\% and 20\%, respectively.

%\paragraph*{Observations}
We have monitored PSR~J1906$+$0746 from 2012-18 with the 305-m William
E.~Gordon Arecibo radio telescope using the Puerto Rico Ultimate
Pulsar Processing Instrument (PUPPI) tuned at a central frequency of
1.38\,GHz. We supplement those observations with archival data from
the Nan\c cay and Arecibo telescopes recorded between 2005 and
2009 \cite{vks+15,sup}. In total, our dataset
comprises 47 epochs spanning from July 2005 to June 2018.
%, about 8\% of
%the spin-precession period of this pulsar.

A previously observed trend of changing separation between MP and IP
profiles with time \cite{vks+15}, has continued. Our data shows that
the slope of the PA under the MP gradually flattens with time, while
the MP becomes simultaneously weaker and subsequently undetectable
towards the end of 2016. This indicates that our LOS has moved out of
the MP emission beam.

The progressive steepening of the PA curve under the IP eventually
leads to a flip of the PA around May 2014 (although it appears flat
due to the 180$^\circ$ ambiguity in the PA), followed by a
flattening. This behaviour is in agreement with the RVM for a pole
crossing our LOS.  The observed change in
the PA curve provides the evidence that the PA swing has a geometrical
origin linked to the magnetic field.

\begin{figure}
\includegraphics[width=\columnwidth]{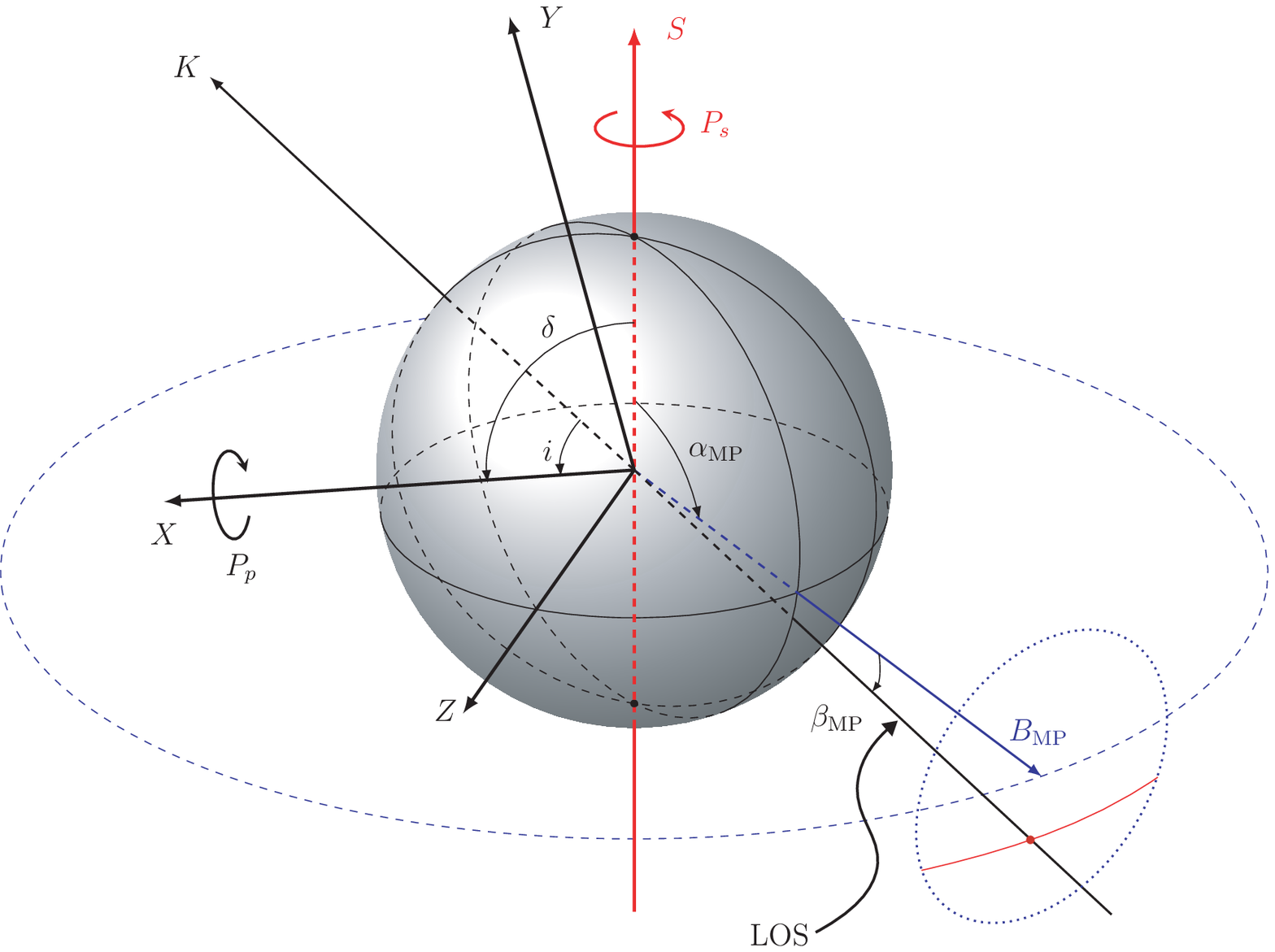}
\caption{{\bf Geometry of PSR J1906$+$0746.} The pulsar rotates with a
  spin period $P_\textrm{S}=144\,\textrm{ms}$ around its spin vector
  ${ S}$ shown with the vertical red arrow. ${ S}$ precesses with a
  period of $P_\textrm{p}=360^\circ /\Omega_{p}\sim 160\,\textrm{yr}$ around
  the total angular momentum vector (misaligned by the angle $\delta$
  from ${ S}$) that can be approximated by the orbital momentum vector
  ${ X}$, perpendicular to the $Y-Z$ orbital plane. $i$ is
  the orbital inclination angle. As the pulsar spins, its magnetic
  pole corresponding to the MP emission, $B_{\textrm{MP}}$, and
  inclined with an angle $\alpha_{\textrm{MP}}$ sweeps the sky along
  the dashed blue trajectory. The MP beam, with the extent
  pictured by the dotted blue circle, crosses our LOS represented by
  the $K$ vector if $|\beta_{\textrm{MP}}| < 22^\circ$. This allows us to
  observe a cut, shown with the red curve, through the MP beam. After
  half a rotation of the pulsar, our LOS can also potentially cut
  through the IP beam if $|\beta_{\textrm{IP}}| < 22^\circ$. For
  clarity, only the MP beam is shown.}
\label{fig:1906_geometry}
\end{figure}

%\paragraph*{Modelling the relativistic spin-precession}
The presence of MP and IP emission, covering a wide range of pulse
longitudes, allows a precise determination of the viewing geometry as
a function of time. We performed a simultaneous modelling of the
polarimetric profiles using the RVM including the effects of
relativistic spin-precession \cite{kw09}, hereafter referred to as the
`precessional RVM'. A total of 53 parameters are included in this
model (see Table~\ref{tab:globalRVM}); the main parameters are the
angle between the rotation axis and the magnetic axis of the MP,
$\alpha_{\textrm{MP}}$, the misalignment angle, $\delta$, the
inclination angle, $i$, and the precession rate, $\Omega_\textrm{p}$.
The phase of the inflection point of the RVM under the MP,
$\phi_{0_{\textrm{MP},k}}$ for each epoch $k$ was also included in the
analysis as a free parameter. The geometry of the system is
illustrated in Fig~\ref{fig:1906_geometry}. We used the Bayesian
sampling tool \textsc{PolyChord} \cite{hhl15} to explore the parameter
space of the model \cite{sup}. The 1-$\sigma$ uncertainty intervals
were derived from the one-dimensional marginalized posterior
distributions (shown in Fig.~\ref{fig:globalRVM}), giving the 68 per
cent confidence level on each parameter.

%on the posteriors give the 68 per cent confidence level on each
%parameter quoted here.

We measure $\alpha_{\textrm{MP}}=99^{\circ}.41\pm0^{\circ}.17$ (and
therefore $\alpha_{\textrm{IP}} = 180^{\circ} - \alpha_{\textrm{MP}} =
80^{\circ}.62\pm0^{\circ}.17$), which combined with a separation
between the MP and IP close to 180$^{\circ}$ confirms that
PSR~J1906$+$0746 is an orthogonal rotator and that the MP and IP
emissions originate from opposite magnetic poles.  The large
$\delta=104^{\circ}\pm9^{\circ}$ is consistent with the pulsar having
formed in an asymmetric supernova (SN) explosion, producing a tilt of
the spin axis of the younger pulsar in the binary pair \cite{vks+15}
relative to the pre-SN orbit \cite{jan17}.  As the spin vector
contribution to the total angular momentum vector is negligible, we
also interpret $\delta$ as the angle between the orbital angular
momentum vector and the pulsar spin vector. With $\delta \sim
104^{\circ}$, this suggests that the pulsar spin axis  lies close to
the orbital plane of this binary system.

\begin{figure}
\includegraphics[width=\columnwidth,angle=-90]{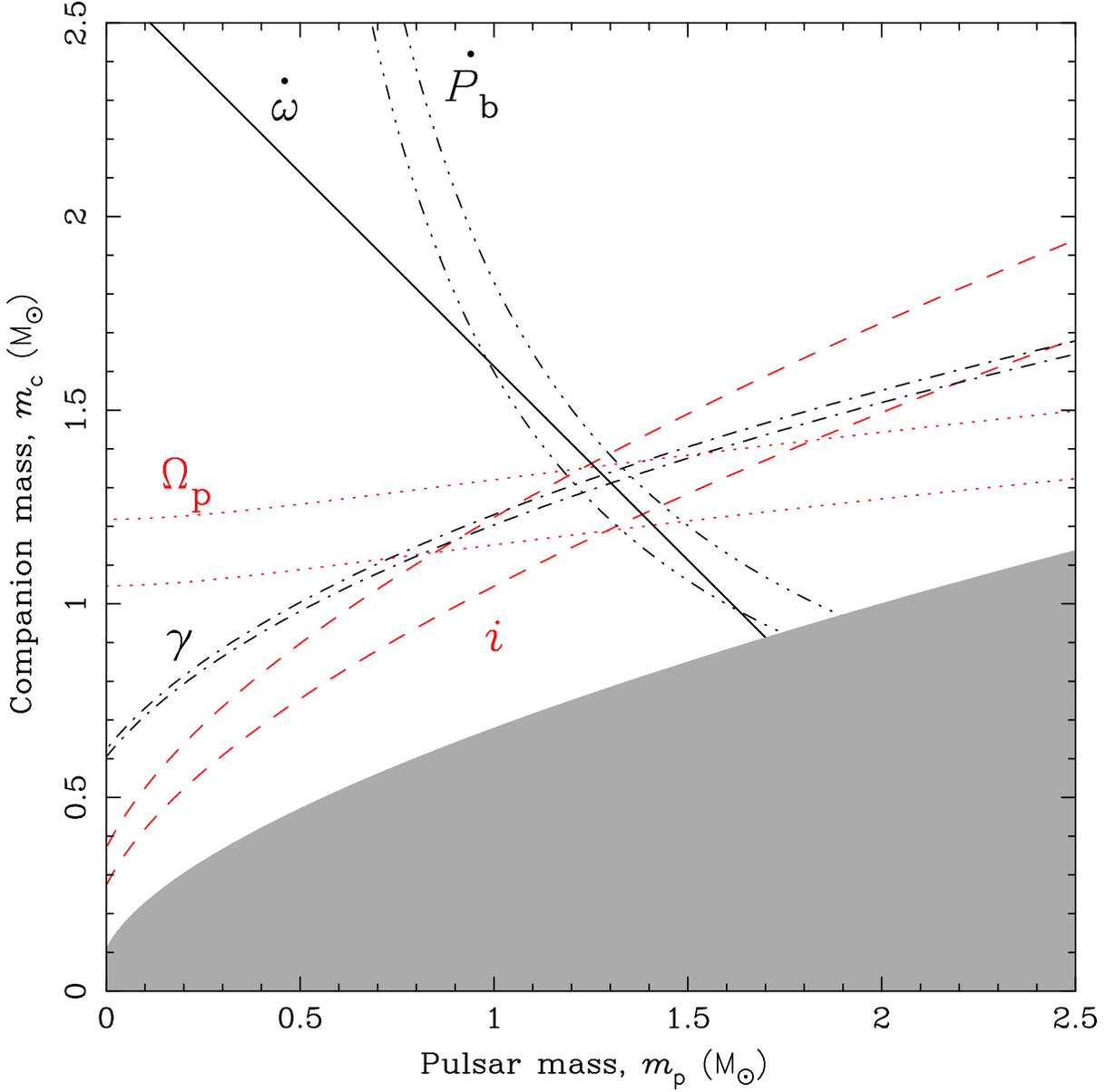}
\caption{{\bf Mass-mass diagram.} The black lines delimit the
  1$-\sigma$ contours from the measurements of the orbital period
  decay, $\dot{P_\text{b}}$, the periastron advance, $\dot{\omega}$,
  and the time dilation, $\gamma$ \cite{vks+15}. The red dotted and
  dashed lines delimit our two additional constraints, the
  measurements of spin-precession rate $\Omega_\text{p}$ and
  inclination angle $i$, respectively. The grey parameter space is
  excluded by the mass function due to $\sin i \leq 1$. All parameters
  are consistent with $m_{\textrm{p}}=1.29\pm0.01$ ~M$_{\odot}$ and
  $m_{\textrm{c}}=1.32\pm0.01$~M$_{\odot}$. }

\label{fig:mm1906}
\end{figure}

This analysis independently determines
$\Omega_\textrm{p}=2.17\pm0.11$\,deg\,yr$^{-1}$ and
$i=45^{\circ}\pm3^{\circ}$ in addition to the three previously
measured PK parameters \cite{vks+15}. This allows us to perform two
additional self-consistent tests of GR in this system (see
Fig.~\ref{fig:mm1906}). Both values agree with the GR predictions
within 1$\sigma$. The constraint on $\Omega_\textrm{p}$ tests GR at a
5\% uncertainty level, tighter than the precession rate measurement in
the Double Pulsar system \cite{bkk+08}. The parametrization of our
model allows us to determine $i$ without ambiguity, in contrast to
information obtained solely from pulsar timing where only $\sin i$ can
be derived.

From our fitted model, we can describe the evolution of the impact
parameters for both the MP and IP, $\beta_{\textrm{MP}}(t)$ and
$\beta_{\textrm{IP}}(t)$ (see Fig.~\ref{fig:precess}) and the
latitudinal extent of the pulsar beam (i.e. the largest value of
$|\beta|$ for which emission is observed).  Our model predicts that in
1998 $\beta_{\textrm{MP}} \sim 5^\circ$ and $\beta_{\textrm{IP}} >
22^\circ$, coinciding with the strong detection of the MP and the
non-detection of the IP in the PMPS data \cite{lsf+06}.  The
appearance of the IP between 1998 and 2004 corresponds to
$\beta_{\textrm{IP}}(2001) \sim +20^{\circ}$. By 2018, our LOS to the
IP had traversed to $\beta_{\textrm{IP}}(2018) \sim -6^{\circ}$ so we
infer a latitudinal extent for the IP of $\sim 20^{\circ}$.  For the
MP, we are only able to map the Southern part of the emission beam,
$-5^\circ<\beta_{\textrm{MP}} \lesssim -22^\circ$, after which the
emission is no longer detected. The appearance of the IP and the
disappearance of the MP at the same $|\beta|$, suggests that the MP
and IP share the same latitudinal extent of $\sim 22^{\circ}$. We
therefore predict that the IP will disappear from our LOS around 2028
to reappear between 2070 to 2090 (Fig. \ref{fig:precess}). The MP
should reappear around 2085--2105.

\begin{figure}
\includegraphics[height=\columnwidth,angle=-90]{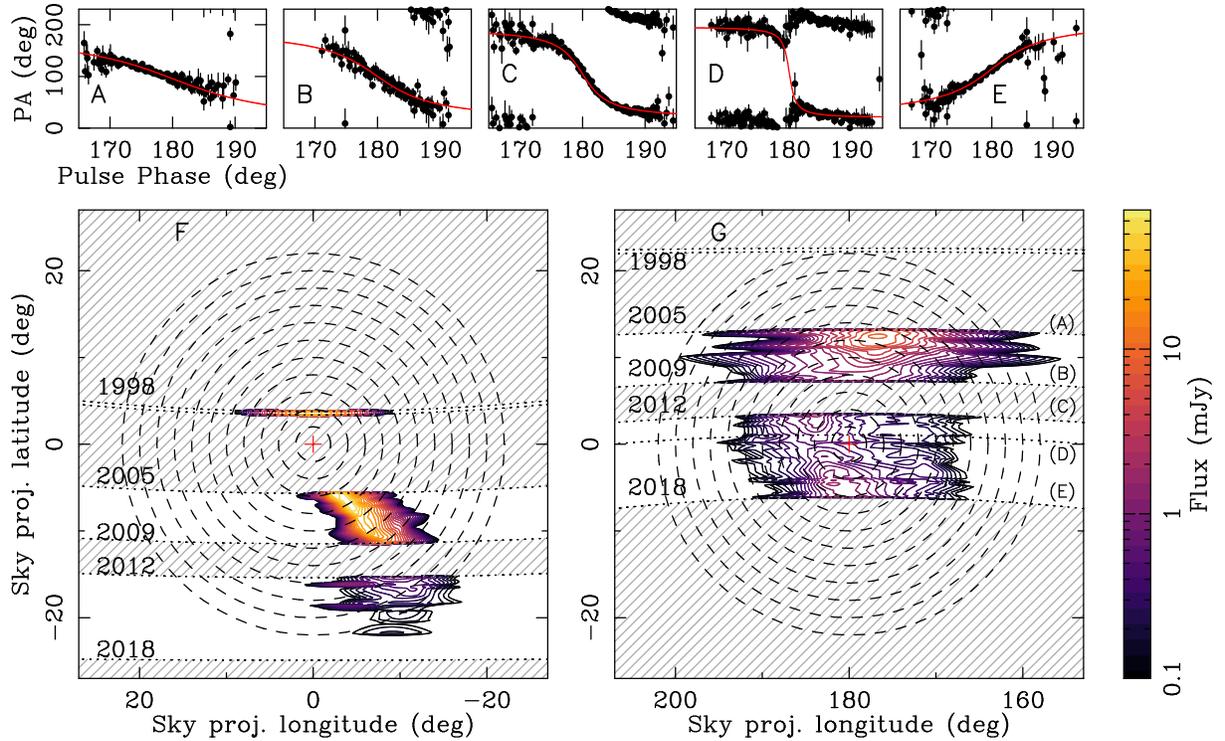}
\caption{{\bf Beam maps of the radio emission.} Two-dimensional
  contours of the reconstructed Stokes I emission maps (on a
  logarithmic color scale), as projected on the sky, for the MP (F)
  and the IP (G). The reference frame of these maps is fixed on the
  vertical spin vector $S$ at longitude zero and 180$^{\circ}$
  pointing towards negative latitude. The red crosses at (0,0) and
  (180,0) indicate the pulsar's magnetic pole on the MP and IP maps,
  respectively, and the dashed circles show increments of two degrees
  in the beam map. The dotted lines represent the LOSs at a given
  year, indicated on the left side. The hatched areas correspond to
  the parts of the maps that were not observed. (A-E) show the PA
  measurements (in black) and the fit by the precessional RVM (the red
  curve) corresponding to the LOSs labelled on the IP map.} 
  %For clarity, the vertical extent of the 1998 observation, represented by
  %the gap between the two dotted curves, has been exaggerated.}% As
  %the 1998 PMPS data do not provide polarisation
  %information, we arbitrarily aligned the centroid of the pulse on the
  %MP magnetic meridian.}
\label{fig:beams} 
\end{figure}

\begin{figure}
\includegraphics[width=\columnwidth]{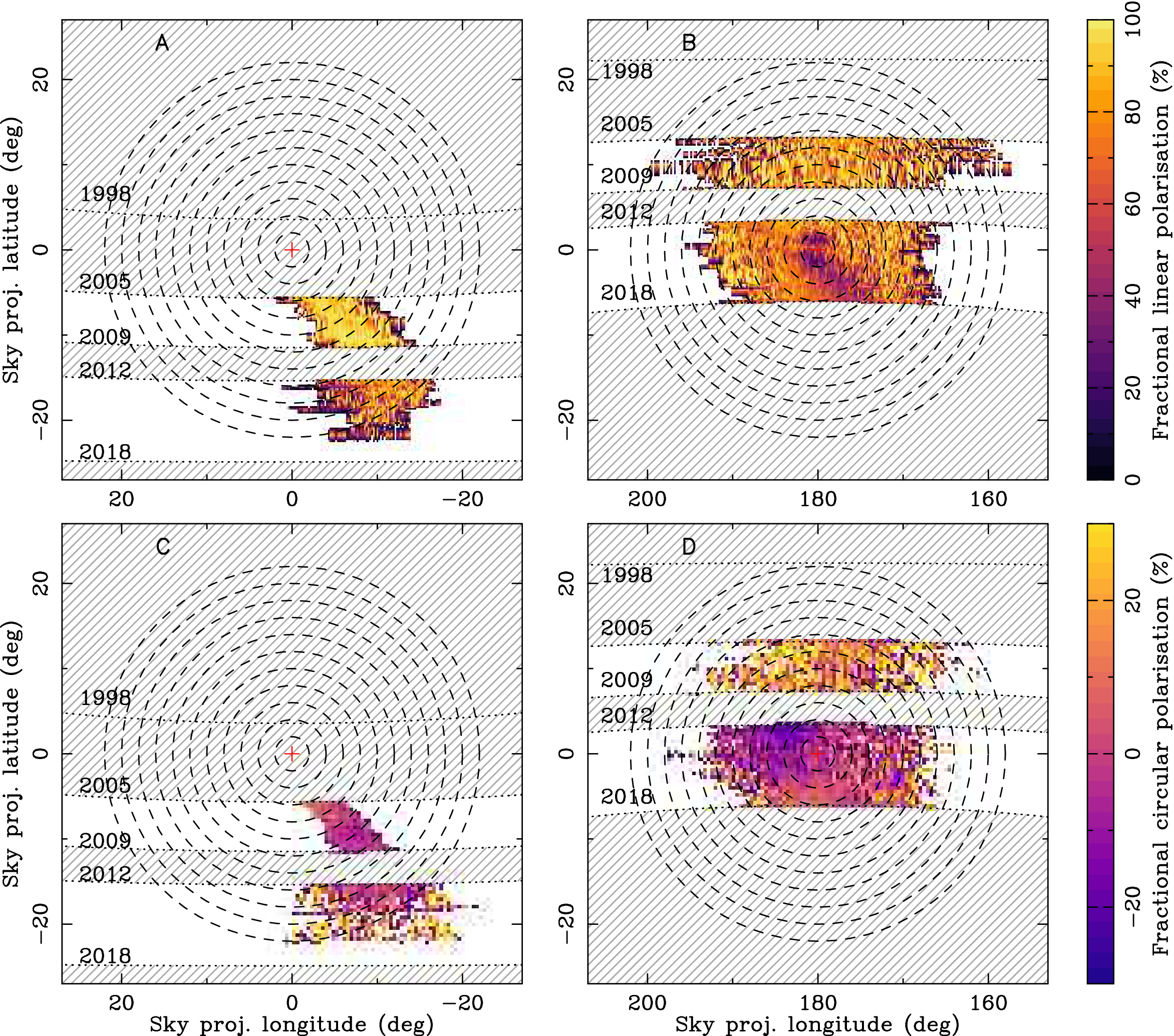} \\
\caption{{\bf Polarisation beam maps.} Same as Figure 3F-G, but
  showing fractional linear polarisation for the MP (A) and IP (B) and
  fractional circular polarisation for the MP (C) and IP (D).}
\label{fig:polar}
\end{figure}

With our determination of the geometry of the system, we can use the
observed pulse profiles recorded at different epochs, including the
reprocessed PMPS data \cite{sup}, to reconstruct \cite{sup} the
sky-projected beam maps of the radio emission from both magnetic poles
(Fig.~\ref{fig:beams}) and its polarisation properties
(Fig.~\ref{fig:polar}). Previous attempts at mapping the pulsar
emission have been published for other pulsars,
e.g. \cite{mks+10}. Such plots must not be viewed as strict maps of
active fieldlines anchored on the polar cap, but rather as projections
of the emission on the sky.

The map shows emission from both sides of the IP magnetic pole, ruling
out models predicting that radio emission is restricted to one side of
the pole (e.g.~\cite{aro83}). The IP emission pattern is not symmetric
in the latitudinal direction or with the MP. The IP weakens when our
LOS crosses the magnetic pole.  This finding matches theoretical
predictions of the current density in the polar cap in case of an
orthogonal rotator, albeit at low radio emission height
\cite{ta13,glp17}.  Radio emission is produced at a height above the
pulsar where any higher-order magnetic field components should have
diminished.  In a standard scenario where emission is produced in the
entire open field-line region, we would expect the radio beam to be
circular. Yet the observed beam is elongated and smaller in the
longitudinal direction than expected (Fig.~\ref{fig:beams}).

% Some
%unpolarised pulse components are shown to appear and disappear on
%timescales of months \cite{sup}.

The passage of the LOS over the IP magnetic pole coincides with a drop
in fractional linear polarisation. At the same time, the circular
polarisation (Stokes V) appears to change sign when crossing the
magnetic pole (Fig.~\ref{fig:polar} and \ref{fig:polar_fraction})
\cite{sup}.  The fractional linear polarisation of the MP decreases as
our LOS moves away from the centre of the MP beam while some unpolarised pulse
components are shown to appear and disappear on timescales of months
\cite{sup}.
% A correlation between PA and a change of sign in V at
%the pulse profile midpoint has previously been deduced from a set of
%pulsars with single LOS observations \cite{rr90} and matches
%theoretical predictions \cite{bp12}.

If we assume that the emission region is symmetric around the magnetic
meridian and that the radiation is beamed in the forward direction of
the relativistic charge flow along the magnetic field lines, then the
rotation of the pulsar magnetosphere in the observer frame causes the
pulse profile to precede the magnetic meridian and the PA to lag it.

Models predict that the total phase shift $\Delta\phi_{\textrm{S}}$
between the midpoint of the pulse profile and the inflection point of
the PA swing is given by $\Delta\phi_{\textrm{S}} \approx 4
h_\textrm{em}/ R_{\textrm{LC}}$, where $h_\textrm{em}$ is the emission
height and the light cylinder radius
$R_{\textrm{LC}}=cP_{\textrm{S}}/2\pi$, with $c$ being the speed of
light \cite{bcw91,dyk08}. This relationship is expected to be valid
for small emission heights, $h_\textrm{em} \lesssim 0.1
R_{\textrm{LC}}$.  But these models predict different contributions to
the aforementioned shifts in the pulse profile and the PA curve. There
is also a shift of the PA in absolute value but this can be neglected
here as $\alpha \sim 90^\circ$ \cite{ha01}.

\begin{figure}
\includegraphics[height=\columnwidth,angle=-90]{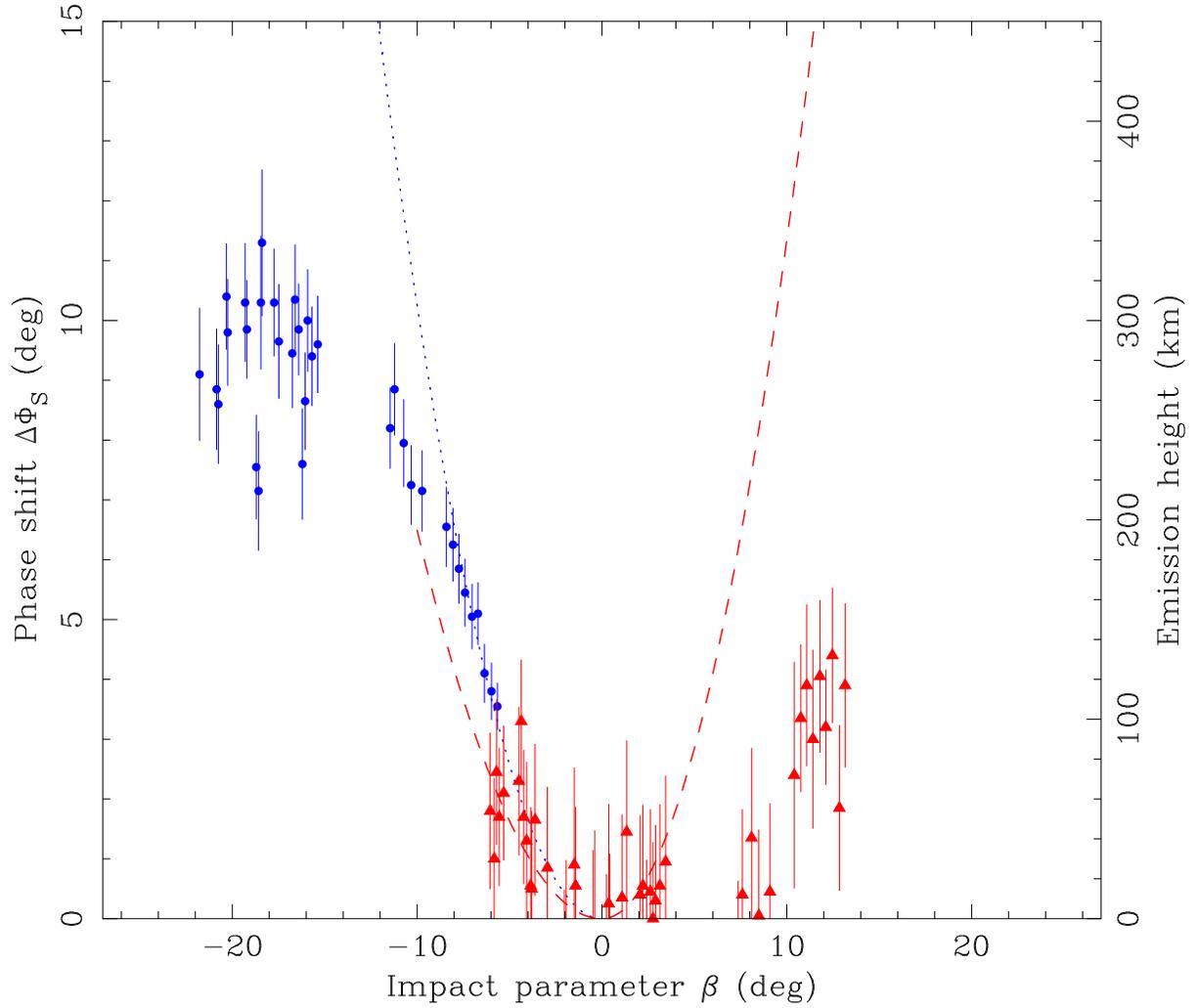}
\caption{{\bf Radio emission heights.}  Phase shift
  $\Delta\Phi_{\textrm{S}}$ of the PA from Stokes I (left axis) with
  the derived emission height (right axis) for the MP (blue data
  points) and IP (red data points) as a function of $\beta$. The blue
  and red dashed curves represent the theoretical emission height as a
  function of $\beta$.  }
\label{fig:height} 
\end{figure}

Measuring the phase shifts $\Delta\phi_{\textrm{S}_{\textrm{MP},k}}$
for the MP at a given epoch $k$ (and its impact parameter,
$\beta(k)$), gives an estimate of the MP emission heights
$h_{\textrm{em}_{\textrm{MP},k}}$, assuming all shifts are due to
rotational effects.  The emission heights for the IP,
$h_{\textrm{em}_{\textrm{IP},k}}$, can be derived in the same way,
also using the RVM inflection points for the IP given by
$\phi_{0_{\textrm{IP},k}} = \phi_{0_{\textrm{MP},k}} +
180^\circ$. These results show that the emission heights range from
close to the surface of the pulsar for low impact parameter $\beta$
(when our LOS is atop of the magnetic pole), to about 320\,km for
large $\beta$, matching the theoretical prediction \cite{ym14}
(Fig.~\ref{fig:height}), even though this theoretical prediction only
depends on $\alpha$ and $\beta$.  Deviations are observed for LOSs
close to the beam edge of the MP, and large $\beta$ for the IP, where
we may observe a different active patch of the IP beam as suggested by
the polarisation properties (e.g. the sign of Stokes V under the IP
has flipped between 2009 and 2012, see Fig.~\ref{fig:polar} and
Fig.~\ref{fig:polar_fraction}).
%A change in relative emission height
%between the MP and IP, and hence, relative phase shift, would partly
%explain the measured change in separation between the midpoint of the
%MP and IP pulse profiles, $\Delta \Phi_{\textrm{I}}$, also shown in
%Fig.~\ref{fig:aberr}.

%In practice, it is difficult to distinguish between a tilted beam
%shape and a gradual shift in phase $\Delta\phi_{\textrm{S}}$ due to
%changing emission heights as a function of $\beta$. If the emission
%height differs for the MP and IP at a given epoch, the difference in
%the respective PA shifts due to corotational effects implies that the
%assumption $\phi_{0_{\textrm{IP},k}} = \phi_{0_{\textrm{MP},k}} +
%180^\circ$ is no longer valid and, therefore, that a single RVM does
%not describe the data optimally. Hence, we also included in the
%precessional RVM analysis a phase offset $\Delta \Phi_{\textrm{RVM}}$
%in the RVM curve between the MP and IP, to allow
%$\phi_{0_{\textrm{IP},k}}$ and $\phi_{0_{\textrm{MP},k}}$ to vary
%independently from each other when the emission from both the MP and
%IP was detected. We find $\Delta \Phi_{\textrm{RVM}}$ is consistent
%with $0^\circ$ within $2\sigma$ for all epochs
%(Fig.~\ref{fig:aberr}). While this confirms the validity of our RVM
%model described above, the only self-consistent description appears to
%be that any corotational phase shift mostly impacts on the pulse
%profile and not on the PA swing.

The latitudinal extent of a pulsar beam determines the pulsar beaming
fraction (i.e.~the portion of sky illuminated by the pulsar
beam). This parameter affects the predicted number of Galactic double
neutron stars (DNSs) population and, hence, the expected gravitational
wave detection rate for neutron star mergers \cite{ok10}.  Usually the
latitudinal extent of the beam cannot be independently determined and
so the expected beam radius $\rho$ for a conal emission beam is
used. This beam shape is usually expected, although
its internal structure and/or filling factor is still a matter of
debate \cite{lk05}.  Based on beam cuts given by pulsar profiles,
various studies (e.g.~\cite{kwj+94}) have suggested $\rho \sim
6.5^{\circ}P_\textrm{S}^{-0.5}$, where the dependence on
$P_\textrm{S}$ is consistent with the expectation for dipolar field
lines.

Previous studies that estimate the DNS population number and merger
rate \cite{ok10} from a sample of pulsars dominated by only two
pulsars (i.e. PSRs J0737$-$3039A and J1906$+$0746) have assumed for
PSR~J1906$+$0746 an uniform radio luminosity across a conal beam of $
\rho\sim 17^\circ$. This value is smaller than our observed
latitudinal extent of $22^\circ$ that combined with
$\alpha_\textrm{MP}=99^{\circ}.4$, gives a large beaming fraction of
0.52 but with substantial luminosity variations.  The simple conal beam
model with uniform luminosity can not be used for PSR J1906$+$0746 to
reliably constrain the DNS merger rate.

%The current estimate for the DNS population and merger rate is
%dominated by only two pulsars, i.e. PSRs J0737$-$3039A and
%J1906$+$0746, with then poorly constrained beaming geometry
%\cite{ok10}. 

%Previous studies that estimate the DNS population number and merger
%rate \cite{ok10} have assumed the conal beam model in the case of
%PSR~J1906+0746 deriving $ \rho\sim 17^\circ$, smaller than our
%observed latitudinal extent of $22^\circ$, and an uniform luminosity
%across the radio beams.  Combining the latitudinal extent of
%$22^\circ$ with $\alpha_\textrm{MP}=99^{\circ}.4$, we derive a large
%beaming fraction of 0.52.
%As the current estimate for the DNS merger rate is dominated by PSRs
%J0737$-$3039A and J1906$+$0746, both with previously poorly
%constrained geometry, more ca.

% The current estimate for tight pulsar-neutron star
%binary mergers come from the similar contribution of the poorly
%constrained geometry of PSRs J0737$-$3039 and J1906$+$0746
%\cite{ok10}.  However the luminosity is not uniform across the beams
%of PSR J1906$+$0746, and careful consideration should be taken regarding
%the beaming fraction when estimating the DNS population and merger
%rate.

\bibliographystyle{Science}
\bibliography{1906}

\iffalse

\fi

\section*{Acknowledgments}
We thank Norbert Wex and Simon Johnston for useful discussions,
Alessandro Ridolfi for sharing the flux calibration observations for
PUPPI and the PALFA collaboration for recording some of the early
Arecibo observations. We acknowledge the use of the Hercules cluster
hosted at the Max Planck Computing and Data Facility in Garching. This
research has made extensive use of NASA's Astrophysics Data System.
{\bf Funding:} GD and MK gratefully acknowledge support from European
Research Council (ERC) Synergy Grant “BlackHoleCam” Grant Agreement
Number 610058. KJL received support from 973 program 2015CB857101,
NSFC U15311243, XDB23010200, 11690024, and funding from Max-Planck
Partner Group.  JvL acknowledges funding from the Netherlands
Organisation for Scientific Research (NWO) under project
"CleanMachine" (614.001.301) and from the ERC under the European
Unions Seventh Framework Programme (FP/2007-2013) / ERC Grant
Agreement n. 617199.  Pulsar research at UBC is supported by an NSERC
Discovery Grant and by the Canadian Institute for Advanced Research.
The Arecibo Observatory is operated by SRI International under a
cooperative agreement with the NSF (AST-1100968), and in alliance with
Ana G. M\'endez-Universidad Metropolitana, and the Universities Space
Research Association.  The Nan\c cay radio Observatory is operated by
the Paris Observatory, associated to the French Centre National de la
Recherche Scientifique (CNRS) and to the Universit\'e d'Orl\'eans.
{\bf Author contributions:} GD and MK led the analysis and the writing
of the manuscript. KL and AJ provided theoretical input and
analysis. JvL, IS and LK performed the Arecibo observations. IC
performed the Nan\c cay observations. All authors commented on the
manuscript.  {\bf Competing interests:} The authors declare no
competing interests. {\bf Data and materials availability:} The Parkes
data are publicly available and can be requested at
https://data.csiro.au/dap/home (scan PM0055\_01341\_21).  The Nan\c
cay and Arecibo datasets can be downloaded from
https://doi.org/10.5281/zenodo.3358819.  The modelRVM tool is
available from https://doi.org/10.5281/zenodo.3265755.
The output parameters of our modelling are listed in Tables S1 and S2.

\section*{Supplementary Materials:}

Materials and Methods

Supplementary Text

Figures S1 to S6

Tables S1 to S2

References (31$-$44)

% 1906 Mass Mass

% Beam plots

% Polar fraction plot

% 1906 Emission height

\newpage

\input{paper-supplementary}

\end{document}

%% file: paper-supplementary.tex
\setcounter{page}{1}

\section*{Materials and Methods}

\beginsupplement

\paragraph*{Calibration and post-processing of the data.}
\label{sec:obs_calib}
PSR J1906$+$0746 was observed with the Nan\c cay Radio Telescope
between 2005 and 2009 with the Berkeley-Orleans-Nan\c cay (BON)
backend at a central frequency of 1.4~GHz \cite{vks+15}. However,
there was at the time no observation of the polarised noise diode
along with the pulsar observation that are typically required to
calibrate polarisation data.

To calibrate in polarisation the Nan\c cay archival data, we applied a
matrix-template matching calibration scheme \cite{van13,bjk+16}.  We
construct the Jones matrix \cite{jon41} by comparing a well-calibrated
observation of a reference pulsar to a non-calibrated observation of
the same pulsar recorded with the same backend at the same observing
frequency.

We choose the millisecond pulsar (MSP) PSR B1937$+$21 as our reference
pulsar due to its well known polarimetric properties \cite{ymv+11},
large flux density and high observing cadence at 1.4\,GHz with the
Nan\c cay radio telescope between 2005 and 2009.  The reference
observation comes from a 2010 observation of this pulsar with the BON
backend that has been calibrated with an observation of the noise
diode. These reference calibrated observations were found to be
consistent with the previously published polarisation results
\cite{ymv+11}.

Each time a change in the instrumental response of the BON backend is
detected through visual inspection of the polarisation data of PSR
B1937$+$21 between 2005 and 2009, a new calibrator solution that
encompass the observations showing the same polarimetric response is
derived. When a reliable calibrator covering a given time span could
not be obtained, the corresponding data were discarded from the
analysis. A total of 29 calibrator solutions were necessary to cover
the five years of BON observations.

As a sanity check, we applied our set of 29 calibrators to the
uncalibrated observations of PSR J1909$-$3744, another regularly
observed and bright MSP, recorded with the BON backend and covering
the same time span as PSR J1906$+$0746.  We inspected the polarisation
profiles and measured the average PA value under the main pulse to
check for any secular variations or artifacts that might have been
introduced by our multi-epochs calibrators without finding any
systematics other than caused by the diurnal variation of the
ionospheric electronic content.

After applying the calibrators to the BON data, we averaged the
observations to form a set of pulse profiles each spanning up to three
months. As both the observing cadence and the signal-to-noise ratio
(S/N) of this pulsar decreased with time, we only made use of 9
averaged profiles spanning June 2005 to July 2007.

PSR~J1906$+$0746 was also observed with Arecibo and the Astronomical
Signal Processor (ASP) pulsar backend between May 2006 and August 2009
at a frequency of 1.44~GHz \cite{vks+15,kas12}. Using observations of
the pulsed noise diode, the pulsar B1929$+$10 and the quasar
B1442$+$101 for flux calibration, we calibrated five high S/N
observations spanning May 2008 to August 2009.

Other archival data were
discarded because they could not provide calibrated polarisation
information or because the S/N of the pulse profiles was too low to
contribute to the analysis while increasing the dimensionality of our
modelling.

In March 2012, we exploited the then newly available PUPPI backend at the
Arecibo radio telescope to resume the monitoring of PSR~J1906$+$0746 at
1.38~GHz, usually observed for 2 hours on a monthly basis. We have 33
calibrated observations spanning March 2012 to June 2018. These new
data allow for polarisation and flux calibration of the recorded
Stokes profiles through observations of a pulsed noise diode and the
quasar PKS 2209$+$080.

We measured the effect of Faraday rotation on Stokes Q
and U of the BON and PUPPI data and found the average rotation measure
(RM) to be $152.3\pm0.7$\,rad\,m$^{-2}$, in agreement with
previous results \cite{lsf+06}. The ASP data were not used to estimate
the RM due to their limited bandwidth but all data were corrected for
Faraday rotation using the average RM. These observations are
summarized in Table~\ref{tab:data}.

\begin{figure}
\includegraphics[height=\textwidth,angle=-90]{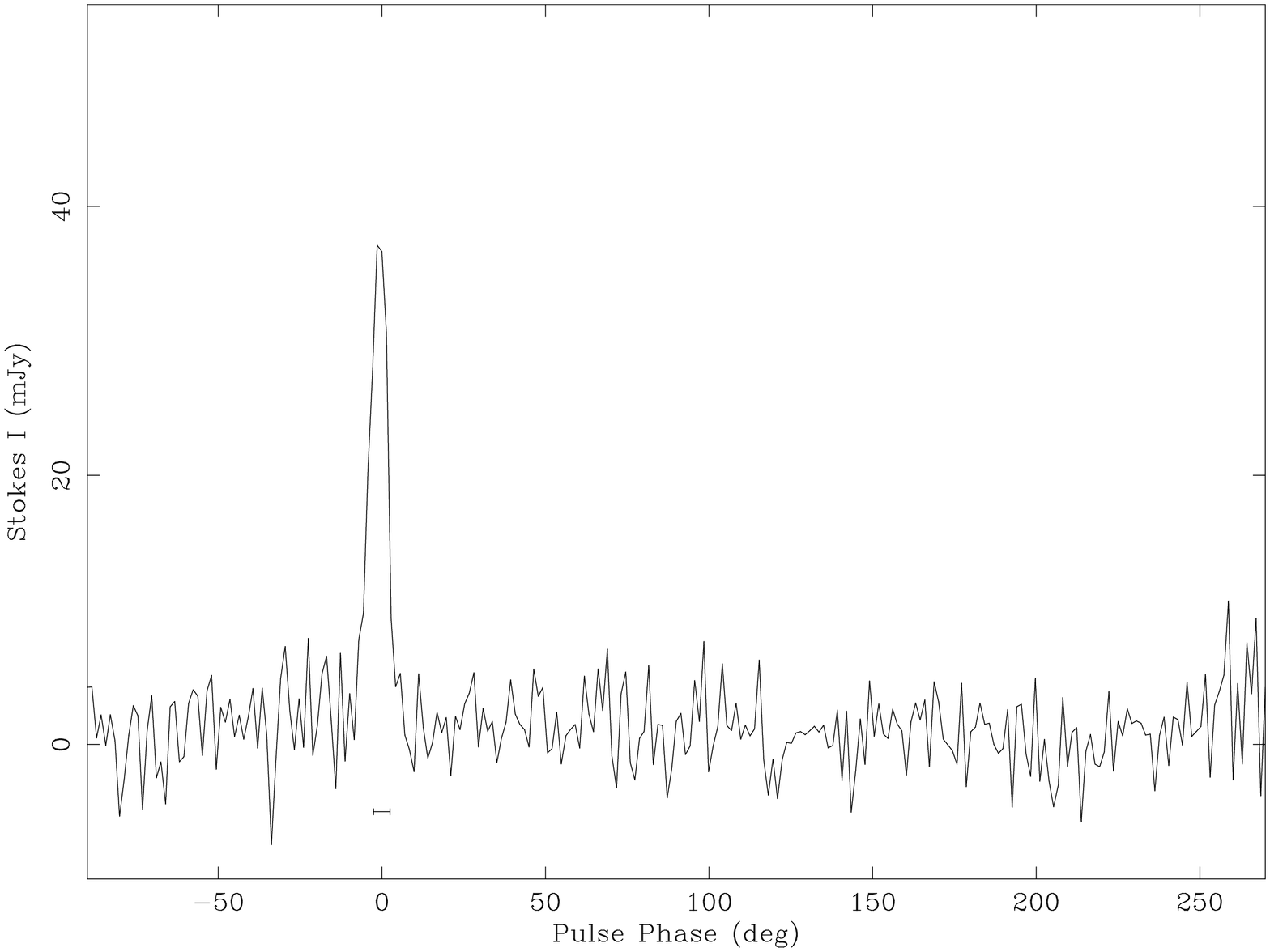}
\caption{{\bf Archival pulse profile from 1998.} Pulse profile of
  PSR~J1906$+$0746 from the PMPS archival data as observed on August
  3rd, 1998. The scaling of the Stokes I in mJy has been done under
  the assumption of the radiometer equation and the PMPS parameters
  (see text). The horizontal error bar under the pulse shows the
  dispersion broadening of the pulse due to the finite channel
  bandwidth of the PMPS data.}
\label{fig:pmps_profile}
\end{figure}

In addition to the data described above, we reprocessed the
PMPS \cite{mlc+01} archival data that show a detection of
PSR~J1906$+$0746 in 1998 \cite{lsf+06}.  The sky position of
PSR~J1906$+$0746 was observed on August 3rd, 1998, i.e. Modified
Julian Date (MJD) 51028, at a central frequency of 1374~MHz and
recorded in 1-bit search-mode filterbank data. Using \textsc{presto}
\cite{ran01}, we dedispersed and folded these time-series data, using
the estimated dispersion measure,
$\textrm{DM}=217\,\textrm{pc}\,\textrm{cm}^{-1}$, and pulse period of
the pulsar, $P_{\textrm{S}}\sim 144.079$\,ms, respectively, to create
an average pulse profile. Due to the 3-MHz wide channels used to
record the PMPS data, intra-channels dispersion,
$\tau_{\textrm{DM}}\sim2\textrm{ms}$, is expected to widen the
observed pulse profile that consists of a single component (see
Fig~\ref{fig:pmps_profile}). This component is assumed to be the MP
\cite{lsf+06}, as later confirmed by our prediction of $\beta$ in
1998, and we set the midpoint of this component to be at longitude
0$^\circ$.  As the PMPS data only provide Stokes I information, we
simply flux calibrated the pulse profile assuming the radiometer
equation (see e.g. \cite{lk05}) and the PMPS parameters
\cite{mlc+01}. The uncertainty of the flux-calibration using the
radiometer equation is assumed to be 20\%. The PMPS pulse profile is
therefore only presented in the Stokes I sky-projected beam maps to
constrain the latitudinal extent of the beam emission.

\paragraph*{RVM fitting with \textsc{modelRVM}}

To fit the PA of the linear polarisation to the RVM, we follow the
International Astronomical Union convention that measures the PA
counter-clockwise on the sky. We also make use of the technique
introduced in \textsc{psrmodel} as part of the \textsc{psrchive}
package \cite{vdo12} that models the observed linear polarisation
${L}$ as a complex quantity, ${L} = {Q}+ i{U}$, given the measured
Stokes parameters ${Q}$ and ${U}$.  In \textsc{psrmodel}, the phase of
${L}$, i.e. the PA $\Psi$, is predicted by the RVM while the amplitude
of ${L}$ is a free parameter. Modelling the complex ${L}$ has a number
of advantages \cite{vdo12},
e.g. the errors on ${Q}$ and ${U}$ are normally distributed. However,
the dimensionality of the modelling increases linearly with the number
of data points included.  We derive the
complex RVM model that allows us to fix the amplitude of ${L}$ instead
of setting it free, greatly reducing the dimensionality of the
problem.  This approach has been introduced in \textsc{modelRVM} \cite{des19}.  In
\textsc{modelRVM}, the estimation of the RVM parameters can be done
with traditional $\chi^2$-minimization techniques through the
\textsc{Minuit} library from the \textsc{ROOT} data analysis package
\cite{br97} or using the nested sampling softwares \textsc{MultiNest}
\cite{fhb09} and \textsc{PolyChord} \cite{hhl15}.

The likelihood $\Lambda$ of using the $Q$ and $U$ data is
\begin{equation}
\log \Lambda = -\frac{1}{2} \sum_{k = 1}^{N_{\tmop{bin}}} \left\{_{}
\frac{[ Q_k - \Re(L_k)]^2}{\sigma_{Q_k}^2} + \frac{[ U_k -
    \Im(L_k)]^2}{\sigma_{U_k}^2} \right\} \, + \tmop{constant},
\end{equation}
where $L_k$, $\sigma_{Q_{_k}}$ and $\sigma_{U_k}$ are the modelled
complex linear polarisation intensity and the noise levels of $Q$ and
$U$ in the $k$-th pulsar phase bin, respectively. $\Re(L_k)$ and
$\Im(L_k)$ are the real and imaginary parts of $L_k$,
respectively. $L_k= L_{n,k} e^{2i\Psi_k}$, where the amplitude
$L_{n,k}$ is a free parameter and $\Psi_k$ is predicted by the RVM
given $\alpha, \beta$, the phase of the RVM inflection point
$\phi_{0},$ and the PA value at phase $\phi_{0}, \psi_0$.  $N_{\rm
bin}$ is the number of phase bins under the MP and IP that satisfy the
criteria $L>1.5\sigma_\textrm{N}$, where $\sigma_\textrm{N}$ is the
standard deviation of the total-intensity off-pulse noise.  In total,
we have $2 N_{\rm bin}$ data points and $N_{\tmop{bin}} + 4$ model
parameters.

Now, in order to reduce the dimensionality of the model and search
only for the four parameters of interest, i.e. $\alpha$, $\beta$,
$\phi_{0}$ and $\psi_0$, we marginalize over the
parameters $L_i$, that $\Lambda' = \int \Lambda d L_k$, giving
\begin{equation}
\label{eq:likeli}
\log \Lambda' = -\frac{1}{2} \sum_{k = 1}^{N_{\tmop{bin}}} \left\{_{}
\frac{[ Q_k - \Re(L_k')]^2}{\sigma_{Q_k}^2} + \frac{[ U_k - \Im(L_k')]^2}{\sigma_{U_k}^2} \right\} + \tmop{constant}.
\end{equation}
$L_k'= L_{n,k}'  e^{2i\Psi_k}$ 
%and  $L_{n,i}'$ is determined via $\partial \log \Lambda / \partial
%\tmop{logL}_i |_{L_i = L_i'} \nobracket=0$, giving
and  $L_{n,k}'$ is given by
\begin{equation}
L_{n,k}' = \frac{\frac{Q_k q_k}{\sigma_{Q_k}^2} + \frac{U_k
    u_k}{\sigma_{U_k}^2}}{\frac{q_k^2}{\sigma_{Q_k}^2} +
  \frac{u_k^2}{\sigma_{U_k}^2}},
\end{equation}
where  $q_k = \cos 2\Psi_k$ and $u_k = \sin 2 \Psi_k$.

We used the likelihood $\Lambda'$ in \textsc{modelRVM} to estimate for
all epochs the four RVM parameters sampled from uniform priors with
\textsc{MultiNest}.
 The results
of $\beta_\textrm{MP}$ and $\beta_\textrm{IP}$ derived from the
modelling of the RVM using $\phi_{0_{\textrm{MP}}}$ and
$\phi_{0_{\textrm{IP}}}$, respectively, are shown in
Table~\ref{tab:data}.

\paragraph*{Precessional RVM}
As previously discussed in the literature, we expect the absolute PA
value of PSR J1906$+$0746 to change as a function of time due to the
precession of its spin axis around the total angular momentum vector
\cite{dt92,kw09}.  To model this effect in the `precessional RVM', we
combined the likelihood $\Lambda'$ from Equation~\ref{eq:likeli} with
the equations 2 to 9 from \cite{kw09}.

The required parameters are $\alpha_{\textrm{MP}}$, $\delta$, a
reference precessional phase $\Phi_{0}$ and a reference PA $\Delta
\psi_{0}$ at an epoch $T_0$, chosen in \textsc{modelRVM} to be the
epoch of the first observation in the dataset, $T_0= \textrm{MJD}\, 53571$. For
each epoch $k$, the phase of the inflection point of the RVM under the
MP, $\phi_{0_{\textrm{MP},k}}$ is also a parameter in the model. The
inclination angle $i$ and the precession rate $\Omega_\textrm{p}$ can
either be fixed to the GR values or included as free parameter in the
model. The parameters of the model are summarised in
Table~\ref{tab:globalRVM}.

As small deviations from the RVM are observed for regions where we
expect our line of sight to be extremely close to the magnetic pole,
we excluded from our analysis a phase range of $\sim 3^{\circ}$ atop
of the pole for the 8 epochs recorded between MJD 56500 and 57000. As
these phase ranges also correspond to dips in the observed linear
polarisation, we performed a phase-resolved RM analysis on these data
(averaged over $\sim0.35^{\circ}$ of pulse phase) to check for
potential pulsar magnetospheric effects on the polarisation
(see e.g. \cite{ijw19} ). However no trend were observed within our
uncertainties of order 10-20\,rad\,m$^{-2}$.

\begin{figure}
\includegraphics[width=\textwidth]{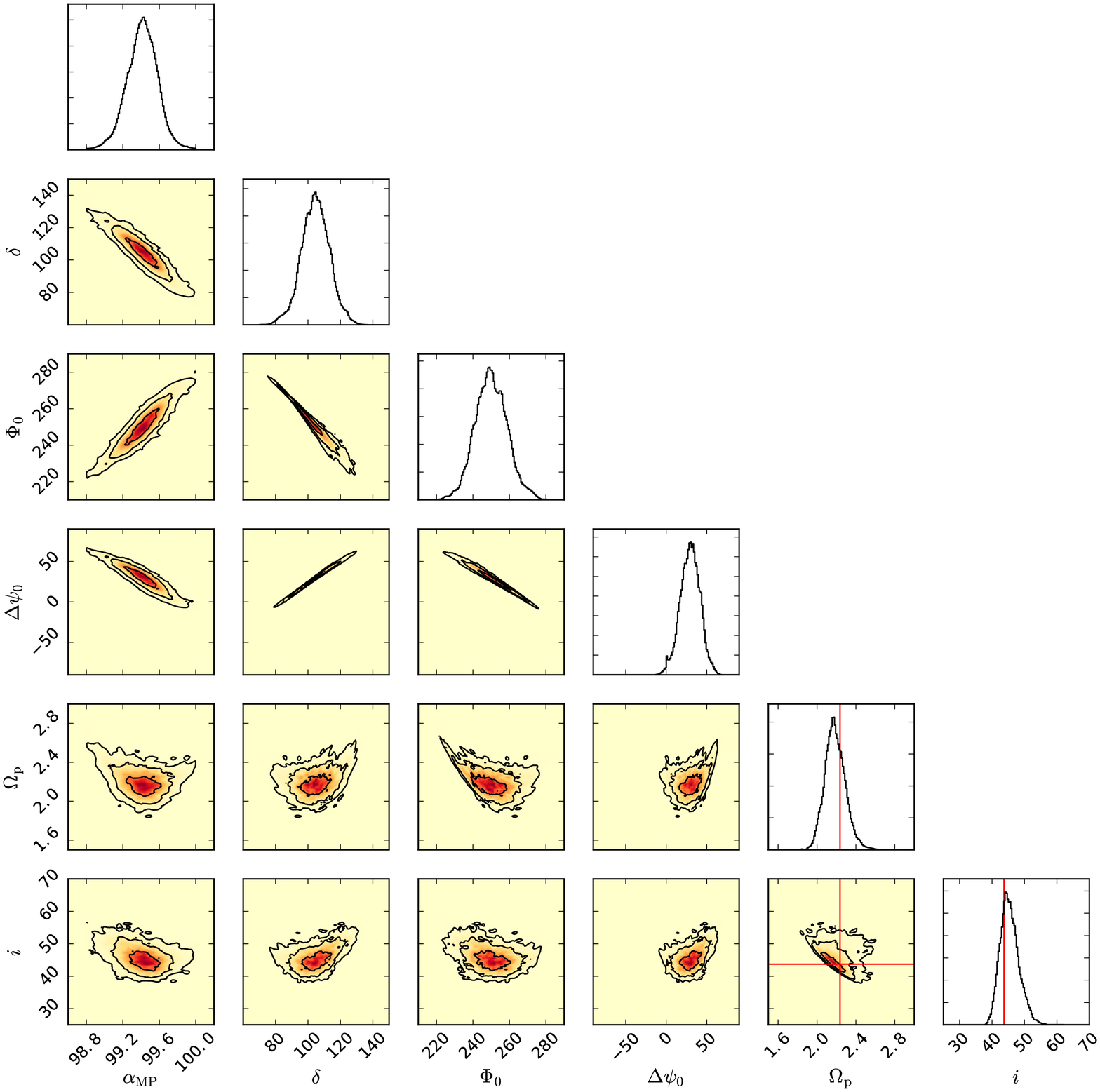}
\caption{{\bf Posterior probability distributions of the precessional RVM.} One and
  two-dimensional marginalized posterior probability distributions
  showing the covariance between the main parameters of interest in
  the precessional RVM. From top to bottom and left to right, the
  represented parameters are the angle between the rotation axis and
  the magnetic axis of the MP, $\alpha_{\textrm{MP}}$, the
  misalignment angle, $\delta$, the reference precessional phase,
  $\Phi_0$, the absolute PA, $\Delta \psi_0$, the precession rate,
  $\Omega_{\textrm{p}}$, and the inclination angle $i$. The red lines
  show the predicted GR values for $\Omega_{\textrm{p}}$ and $i$, in
  agreement with our results.}
\label{fig:globalRVM}
\end{figure}

To account for possible variations in the RM coming from
e.g. ionospheric electron content that can shift the absolute PA
value, we created 2000 data sets where each individual epoch was
corrected for an RM drawn from a normal distribution of width
1.0\,rad\,m$^{-2}$ and centred at 152.3\,rad\,m$^{-2}$.  Then, for
each of the 2000 data sets, we used \textsc{PolyChord} within
\textsc{modelRVM}\ to explore the parameter space of the model with
500 live points, sampled from uniform priors.  We equally
summed the posterior distributions of each parameter from the 2000
data sets to form the final posteriors (Figure~\ref{fig:globalRVM}).

\paragraph*{Reconstructing the beam map}
We begin the beam reconstruction by phase aligning all pulse profiles
with respect to $\phi_{0_{\textrm{MP},k}}$, the inflection point of
the RVM under the MP, to the same phase of 0$^\circ$. Consequently,
the RVM inflection point for the IP is set at the phase of
180$^\circ$. In the case of the PMPS data, we interpret the only
observed component as the MP and therefore arbitrarily set the
midpoint of the pulse to phase $0^\circ$.  We then fit a set of
Gaussian functions to the Stokes I profiles of the MP and IP
separately to form noise-free pulse profiles.  Finally, we separately
perform a two-dimensional spline fit (over time and pulse phase) to
the MP and IP noise-free profiles that we map to the sky-projected
beam grid given the predicted line of sight. The reference frame of
the grid is fixed with respect to the spin vector $S$, vertically
aligned on phase (also labelled longitude on the maps) $0^\circ$ and
180$^\circ$, pointing towards negative latitude. In the maps, at
longitude $0^\circ$ and $180^\circ$, the latitude is equivalent to
$\beta_{\textrm{MP}}$ and $\beta_{\textrm{IP}}$, respectively. We
applied the same two dimensional spline fit to reconstruct the beam
maps for the linear and circular polarisation, but directly on the
linear polarisation $L$ and circular polarisation $V$. 

Because the PMPS data partially suffer from dispersion by the
interstellar medium, the MP emission from the 1998 LOS represented in
the Stokes I emission map (panel (F) in Fig.~\ref{fig:beams}) is
slightly broadened compared to its intrinsic width. For clarity, the
vertical extent of the 1998 observation, represented by the gap
between the two dotted curves, has been exaggerated.

\section*{Supplementary Text}

\begin{figure}
\includegraphics[height=\textwidth,angle=-90]{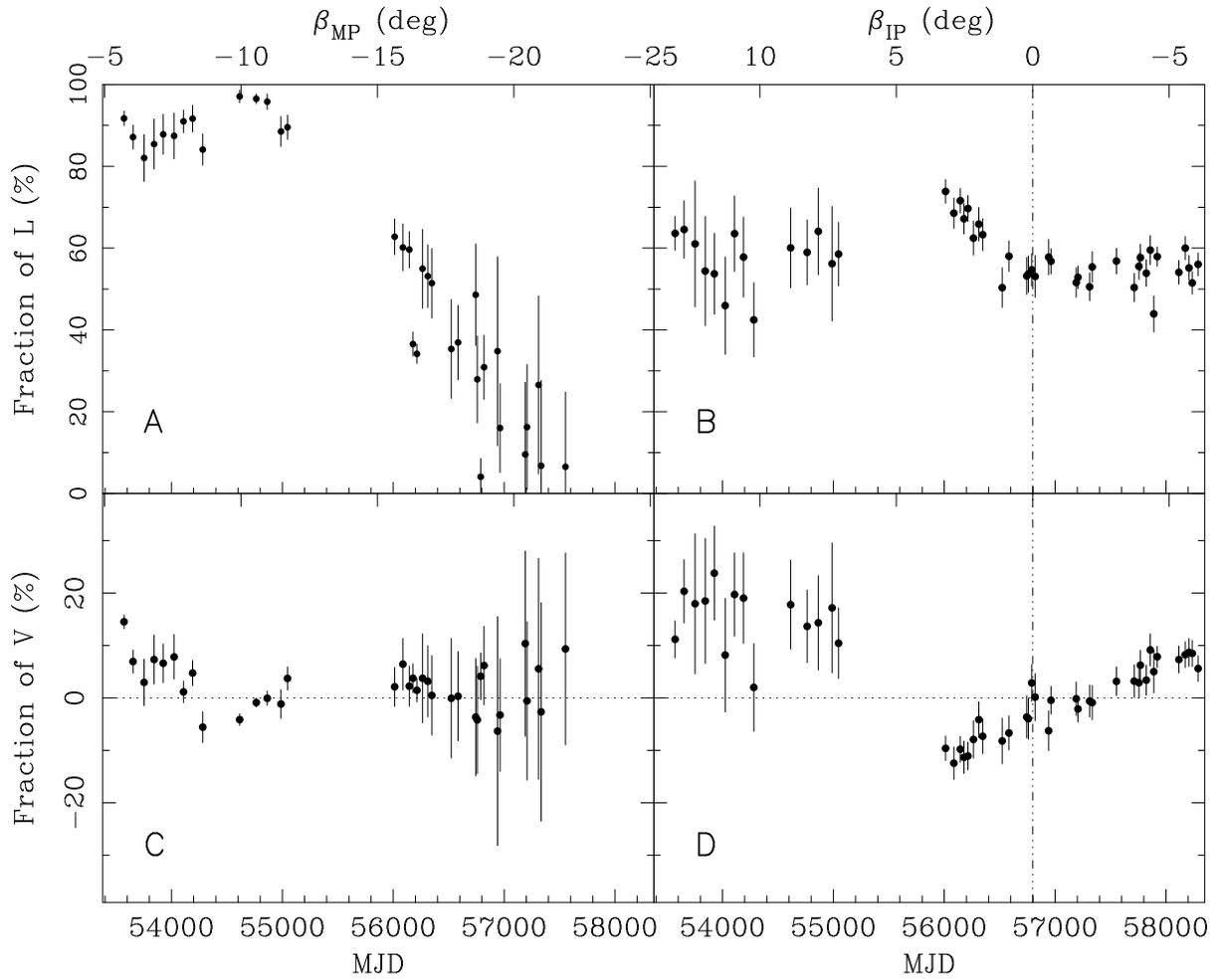}
\caption{{\bf Polarisation fraction of the pulse.} Average percentage
  of linear polarisation for the MP (A) and IP (B) and circular
  polarisation for the MP (C) and the IP (D) as a function of MJD
  (bottom x-axis label) and impact parameter $\beta$ (top x-axis
  label).}
\label{fig:polar_fraction}
\end{figure}

\paragraph*{Polarisation of the beam}
 Stokes V and the gradient of the PA are shown to flip sign as our LOS crosses the IP
magnetic pole (Fig.~\ref{fig:profiles1} and \ref{fig:polar_fraction}). A
correlation between PA and a change of sign in Stokes V at the pulse
profile midpoint has previously been deduced from a set of pulsars
with single LOS observations \cite{rr90} and matches theoretical
predictions \cite{bp12}.

\paragraph*{Caustic emission}
Twisted magnetospheres can cause ``caustics'' from the superposition
of emission originating at different locations in the (e.g. outer)
magnetosphere, as commonly proposed to explain the shape of
light-curves from $\gamma$-ray pulsars (e.g.~\cite{ae98}). At radio
frequencies such caustics should be depolarised \cite{dr03}, and we
speculate that they may be responsible for the strong, unpolarised
pulse component trailing the MP that appeared and disappeared very
rapidly on two occasions (around September 2012, MJD $\sim 56200$,
$\beta_{\textrm{MP}} \sim -16^\circ$, and May 2014, MJD $\sim 56800$,
$\beta_{\textrm{MP}} \sim -18.5^\circ$; see Fig~\ref{fig:profiles1}).

\begin{figure}
\includegraphics[height=\columnwidth,angle=-90]{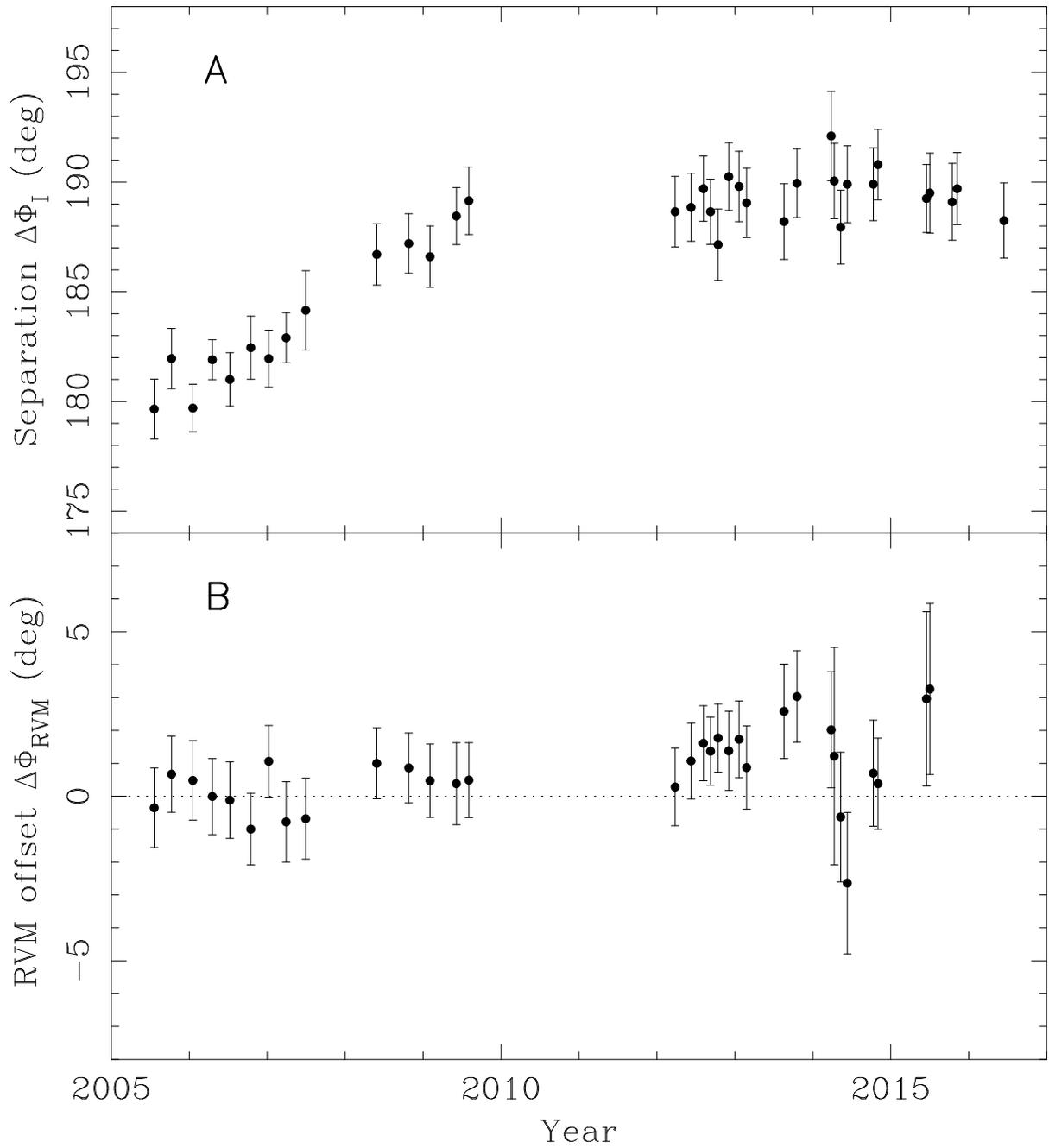}
\caption{{\bf Effects of different emission heights between the MP and IP.}  (A) Separation
  $\Delta \Phi_\textrm{I}$ between the MP and IP Stokes I
  midpoints. (B) Phase offset $\Delta \Phi_\textrm{RVM}$ in the RVM
  curve between the MP and IP. The dashed line at $0^\circ$ shows the
  expected value in case of the simple RVM (i.e. without considering
  rotational effects), or, when considering rotational effects, with
  the emission of the MP and IP originating at the same height.}
\label{fig:aberr}
\end{figure}

\paragraph*{Changes in relative emission height between the MP and IP}
In practice, it is difficult to distinguish between a tilted beam
shape and a gradual shift in phase $\Delta\phi_{\textrm{S}}$ due to
changing emission heights as a function of $\beta$.  A change in
relative emission height between the MP and IP, and hence, relative
phase shift, would partly explain the measured change in separation
between the midpoint of the MP and IP pulse profiles,
$\Delta \Phi_{\textrm{I}}$, also shown in Fig.~\ref{fig:aberr}.  If
the emission height differs for the MP and IP at a given epoch, the
difference in the respective PA shifts due to corotational effects
implies that the assumption $\phi_{0_{\textrm{IP},k}}
= \phi_{0_{\textrm{MP},k}} + 180^\circ$ is no longer valid and,
therefore, that a single RVM does not describe the data
optimally. Hence, we also included in the precessional RVM analysis a
phase offset $\Delta \Phi_{\textrm{RVM}}$ in the RVM curve between the
MP and IP, to allow $\phi_{0_{\textrm{IP},k}}$ and
$\phi_{0_{\textrm{MP},k}}$ to vary independently from each other when
the emission from both the MP and IP was detected. We find
$\Delta \Phi_{\textrm{RVM}}$ is consistent with $0^\circ$ within
$2\sigma$ for all epochs (Fig.~\ref{fig:aberr}). While this confirms
the validity of our RVM model, the only
self-consistent description appears to be that any corotational phase
shift mostly impacts on the pulse profile and not on the PA swing.

\clearpage

\begin{figure}[p]
\includegraphics[width=\textwidth]{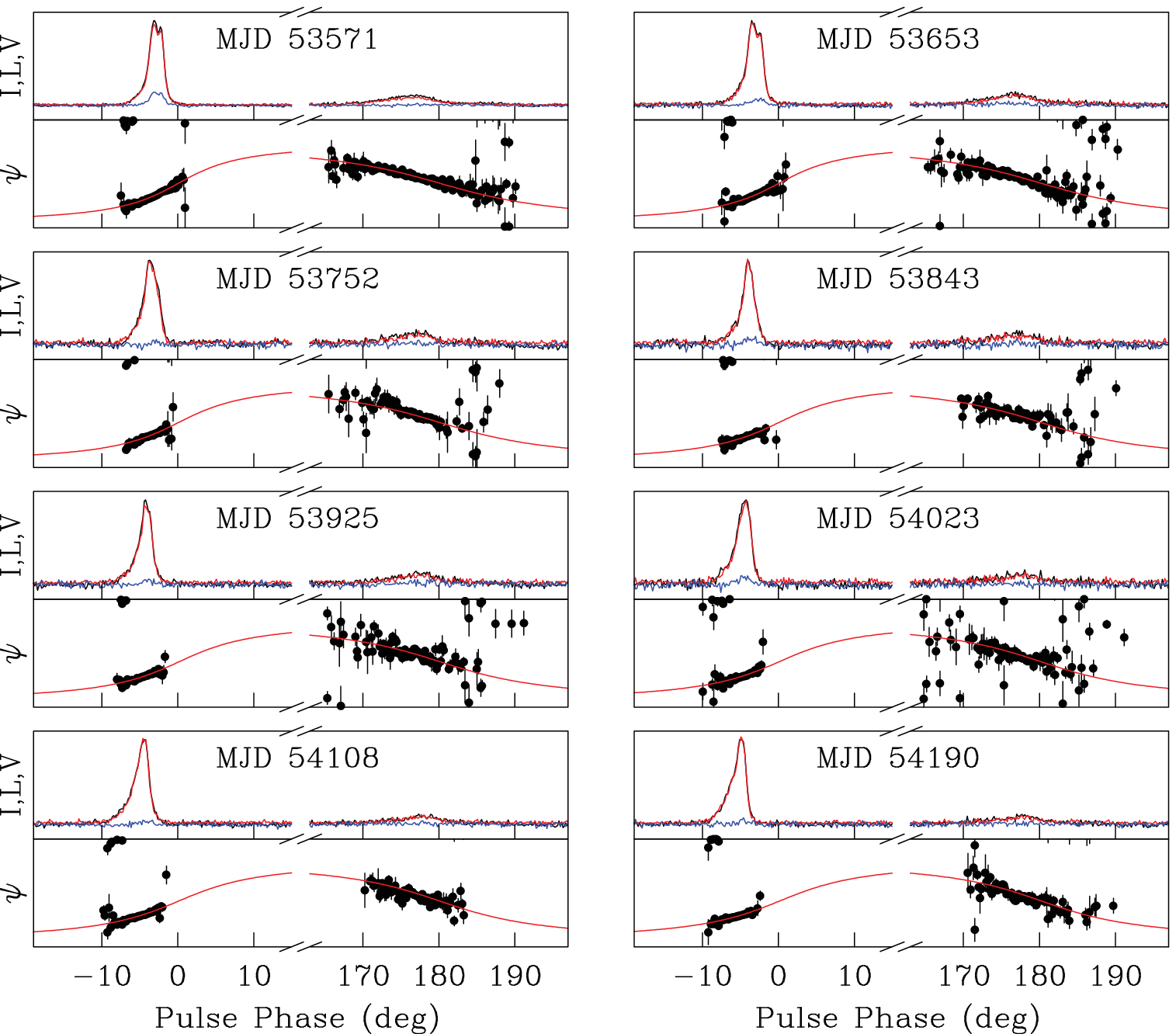}
%\end{figure}
\caption{{\bf Polarisation profiles of PSR~J1906$+$0746.}
  Normalized polarisation profiles and PA measurements of PSR
  J1906$+$0746. To improve readability, only the phase around the MP
  ($\phi_{0_\textrm{MP}}$ at phase 0$^\circ$) and IP
  ($\phi_{0_\textrm{IP}}$ at phase 180$^\circ$) is displayed. For each
  epoch, the normalized total intensity profile I, the linear
  polarisation L and the circular polarisation V are shown in the
  upper panel in black, red and blue colours, respectively. In the
  lower panel with the y-axis ranging from 0 to 230$^\circ$, the black
  data points represent the PA $\psi$ of the linear polarisation and
  the red curve shows the predicted RVM curve from the maximum
  likelihood value of our precessional RVM model. For the 8 profiles
  between MJD 56500 and 57000, the 3$^\circ$-phase ranges atop the IP
  magnetic pole and corresponding to grey PA values were excluded from
  the fit. The PA wraps every 180$^\circ$. The MJD of the averaged
  profiles are written in the upper panels. }
\label{fig:profiles1}
\end{figure}

\begin{figure}
\includegraphics[width=\textwidth]{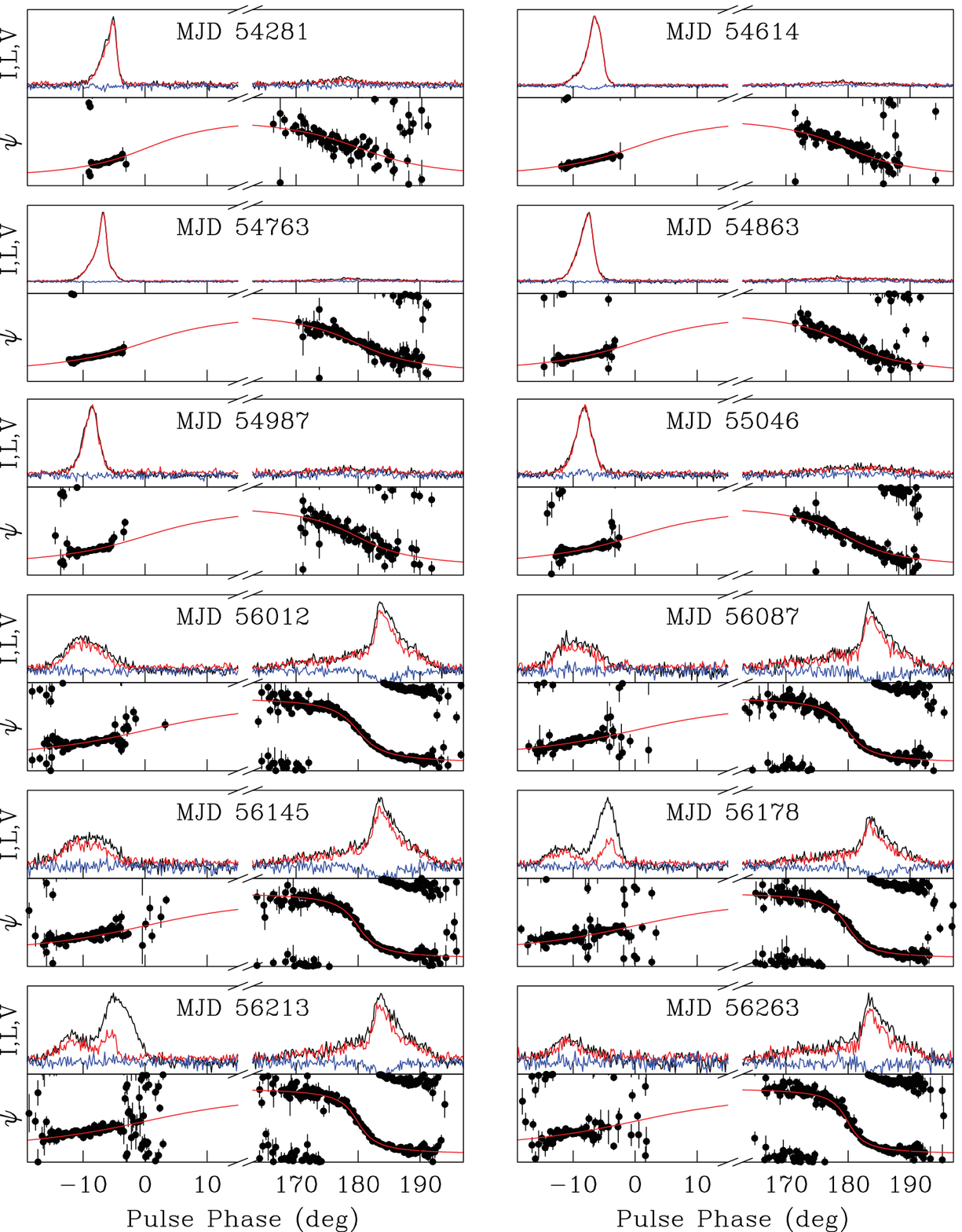}
\contcaption{Continued}
\label{fig:profiles2}
\end{figure}

\begin{figure}
\includegraphics[width=\textwidth]{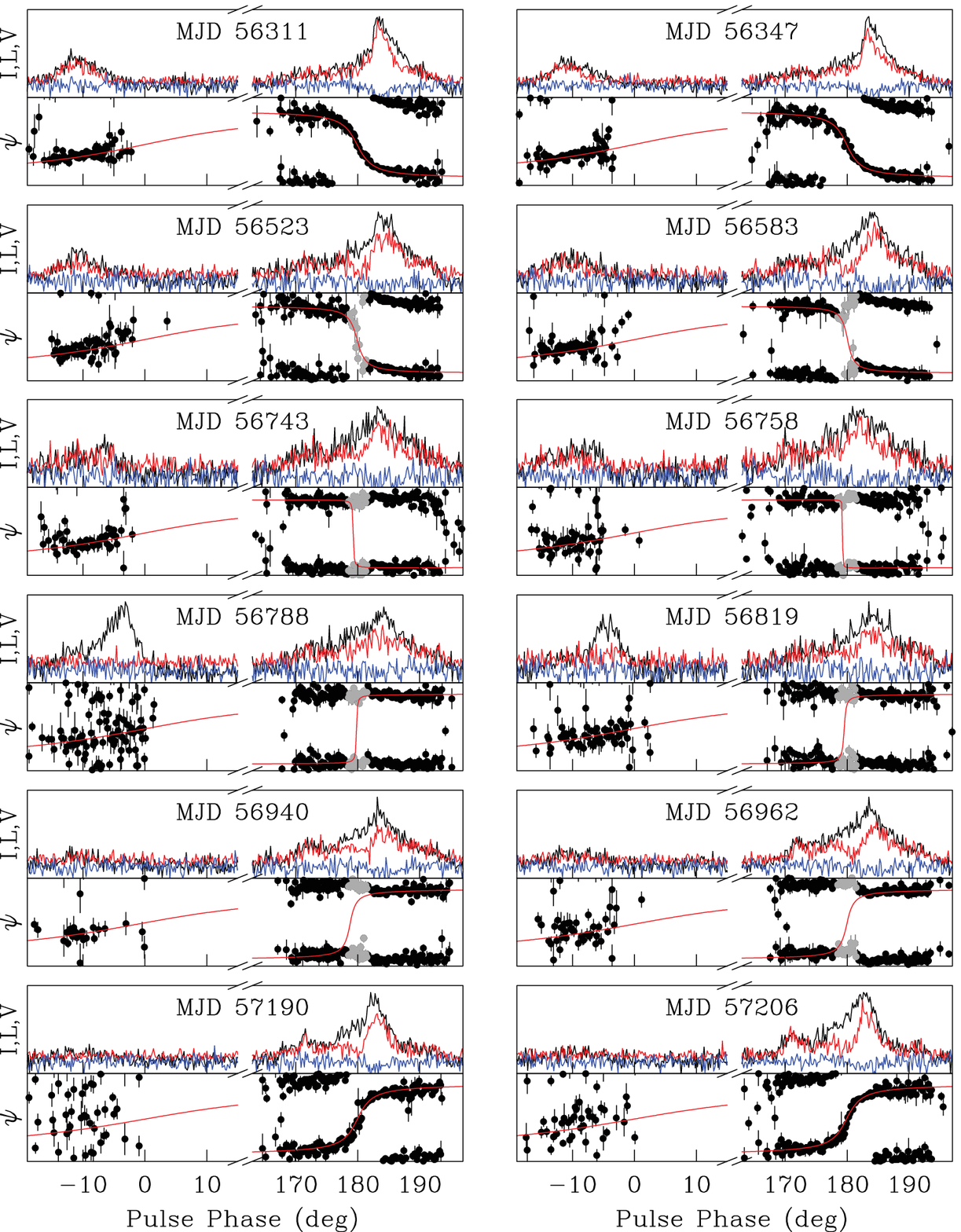}
\contcaption{Continued}
\label{fig:profiles3}
\end{figure}

\begin{figure}
\includegraphics[width=\textwidth]{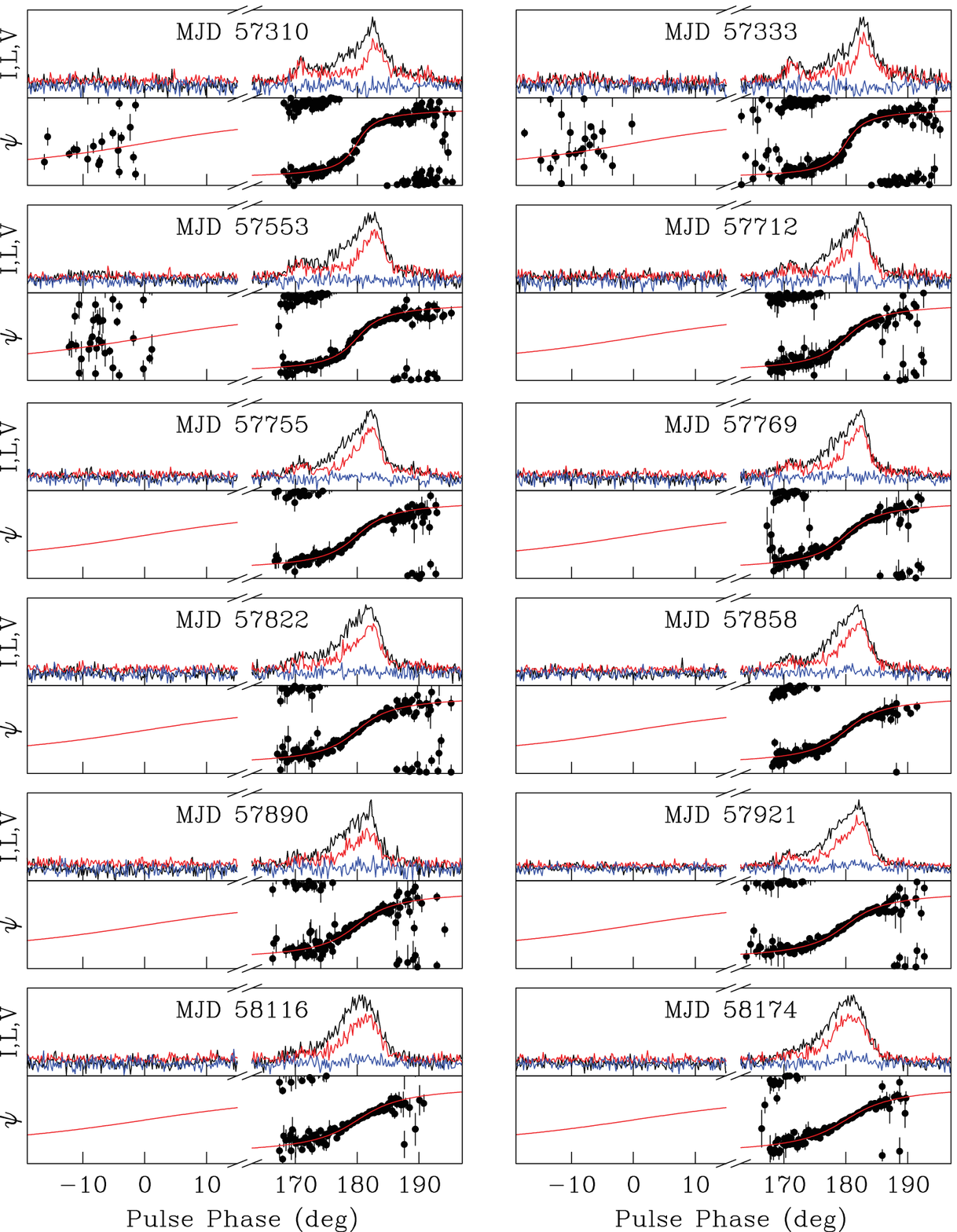}
\contcaption{Continued}
\label{fig:profiles4}
\end{figure}

\begin{figure}
\includegraphics[width=\textwidth]{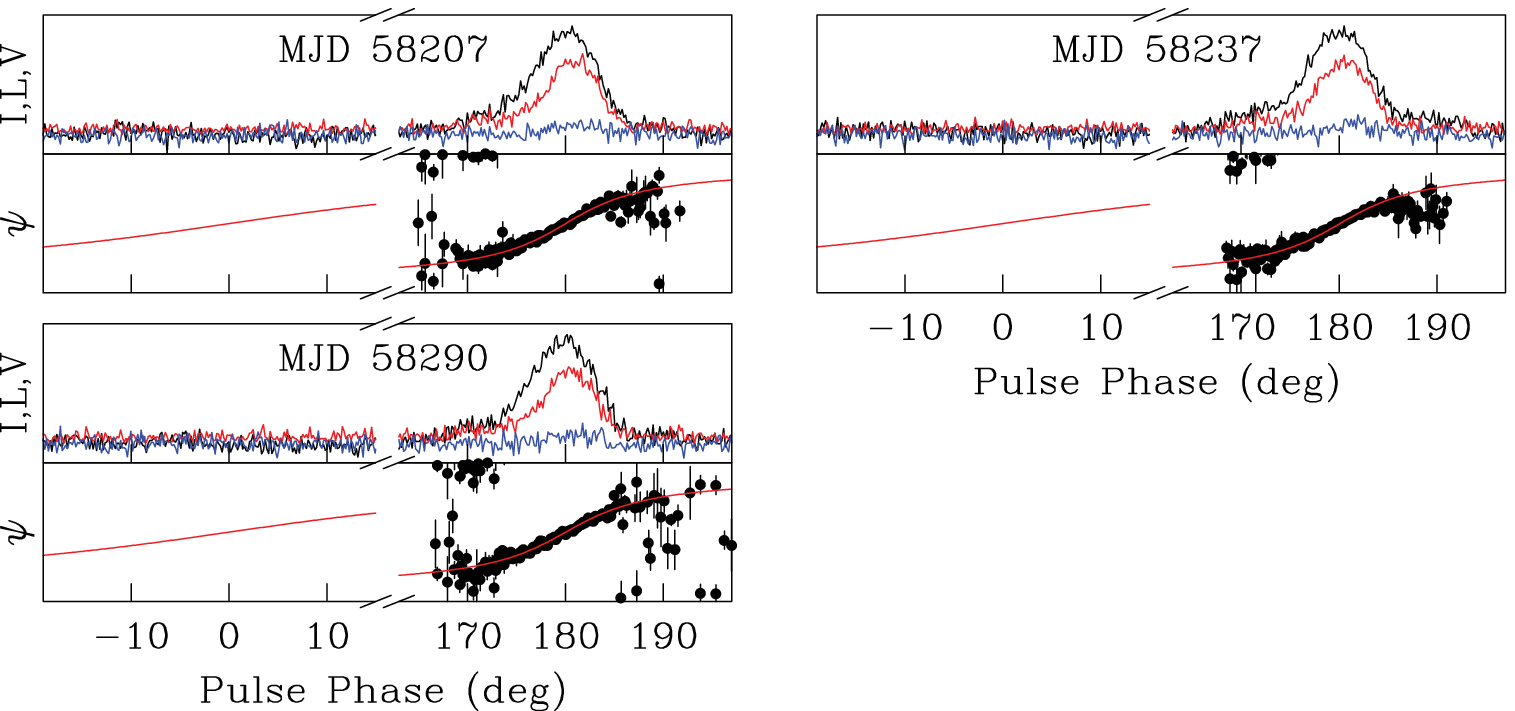}
\contcaption{Concluded}
\label{fig:profiles5}
\end{figure}

\begin{figure}
\includegraphics[height=\textwidth,angle=-90]{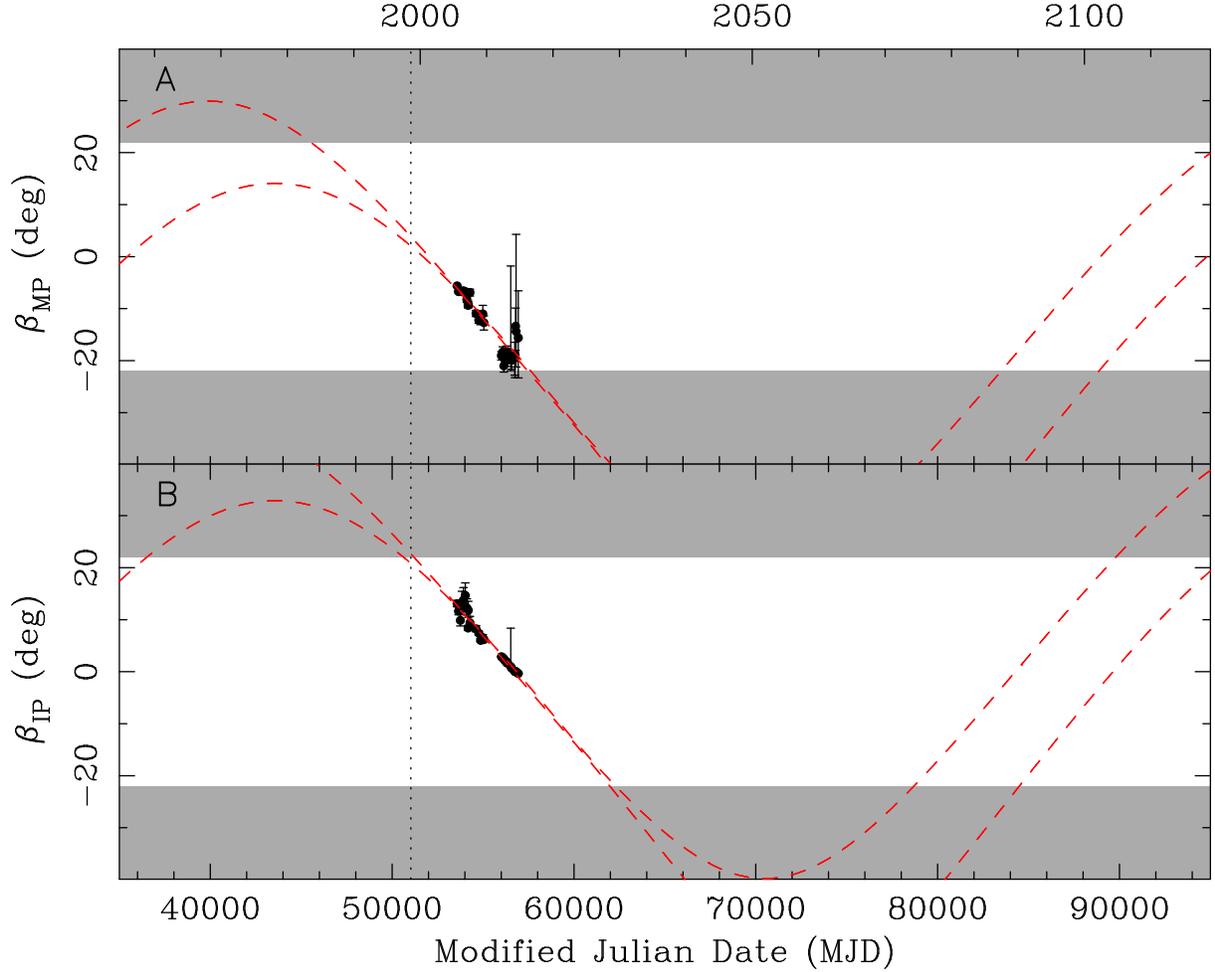}
\caption{{\bf Evolution of the impact parameter as a function of
    time.} (A) and (B) show the secular evolution of
  $\beta_{\textrm{MP}}$ and $\beta_{\textrm{IP}}$, respectively,
  covering the precession period of the pulsar. The data points and
  error bars are taken from the traditional RVM analysis of each epoch
  separately. The red dashed curve delimits the 68 per cent confidence
  levels on $\beta$ as derived from the posterior results. The grey
  areas shows the range of impact parameter, $|\beta| > 22^\circ$,
  where the radio emission is assumed not to be detectable based on
  the observed latitudinal extent of the MP beam. The vertical dotted
  line indicates $\textrm{MJD}=51028$, the epoch of the first
  detection of PSR J1906$+$0746 in the PMPS archival data. At the time
  of the first PMPS detection, our model predicts that
  $\beta_{\textrm{MP}} \sim 5^\circ$ and $\beta_{\textrm{IP}} >
  22^\circ$. }
\label{fig:precess}
\end{figure}

% Average polarisation

% 1906 Emission height                                                                                   

\clearpage

\input{table-params}

\input{table-observations}

%% file: table-params.tex
\begin{table}
\caption{{\bf Parameters of the `precessional RVM'.} Description of
  the parameters used in the `precessional RVM', with the unit
  indicated in parenthesis. The values for
  $\phi_{0_{\textrm{MP},k}}$ with $k \in \{1..47\}$ are not
  reported here as they are considered as ad-hoc alignment offsets in this
  model. }
\label{tab:globalRVM}
\begin{center}
\begin{tabular}{rrl}
\hline
Parameter & Value & Description\\  
\hline
$\alpha_\textrm{MP}$ & $99.41\pm0.17$ & Angle between the spin axis and the MP magnetic pole ($^\circ$) \\
$\delta$ & $104\pm9$ & Angle between the pulsar spin axis\\
 & &  and the total angular momentum vector ($^\circ$) \\
$\Phi_{0}$ & $249\pm10$ & Reference precessional phase at epoch $T_0$ ($^\circ$) \\
$\Delta \psi_{0}$ & $29\pm12$ &  Reference PA value at epoch $T_0$ ($^\circ$) \\
$\Omega_\text{p} $ &  $2.17\pm0.11$ & Precession rate, can be set free or fixed to the value predicted by GR ($^\circ\,\textrm{yr}^{-1}$) \\
$i$ & $ 45\pm3$ & Inclination angle, can be set free or fixed to the value predicted by GR ($^\circ$) \\
$\phi_{0_{\textrm{MP},k}}$ & --- & Phase of the RVM inflection point of the MP for each epoch $k \in \{1..47\}$ ($^\circ$) \\
\hline
\end{tabular}
\end{center}
\end{table}

%% file: table-observations.tex
\begin{longtable}{lcrrSS}
\caption{{\bf Summary of the observations.} Summary of the 47 epochs
  in our dataset. The columns show the MJD of the observation, backend
  name, integration time, estimated RM, impact parameter
  of the MP and IP as given by the individual RVM analysis,
  respectively. The MJD indicated for the BON data corresponds to the
  weighted MJD of the BON observations averaged together over a three
  month span. The RM of the ASP data is not reported here as the
  bandwidth of these data (16~MHz) is too small to derive a meaningful
  value. $\beta_{\textrm{MP}}$ could not be reliably measured after
  MJD 56950 as the MP weakened until disappearance around MJD
  57600-57700.}
\label{tab:data}
\\
\hline
%PSR JName   & N$_{j}$   & $A_{\text{RN}}$ & $\gamma_{\text{RN}}$ & $A_{\text{DM}}$  & $\gamma_{\text{D\M}}$ & Observatory & Frequency & Year Range  & N$_{\text{TOAs}}$ & $E_f$ & $E_q$ \\
\multicolumn{1}{c}{MJD} &
\multicolumn{1}{c}{Backend} &
T$_\textrm{int}$ &
\multicolumn{1}{c}{RM} &
\multicolumn{1}{c}{$\beta_{\textrm{MP}}$} &
\multicolumn{1}{c}{$\beta_{\textrm{IP}}$} \\
&
&
(hr) &
\multicolumn{1}{c}{(rad m$^{-2}$)} &
\multicolumn{1}{c}{(deg)} &
\multicolumn{1}{c}{(deg)}\\
\hline
\endfirsthead
%\multicolumn{12}{crrrrrcccccc}%                                                                       
\multicolumn{6}{c}%                                                                                   
{\tablename\ \thetable\ -- \textit{Continued from previous page}} \\
\hline
\multicolumn{1}{c}{MJD} &
\multicolumn{1}{c}{Backend} &
T$_\textrm{int}$ &
\multicolumn{1}{c}{RM} &
\multicolumn{1}{c}{$\beta_{\textrm{MP}}$} &
\multicolumn{1}{c}{$\beta_{\textrm{IP}}$} \\
 & 
 &
(hr) &
\multicolumn{1}{c}{(rad m$^{-2}$)} &
\multicolumn{1}{c}{(deg)} &
\multicolumn{1}{c}{(deg)}\\
\hline
\endhead
\hline
 53571.9 & BON & 21.5 & $155\pm7$ & ${-5.6_{-0.1}^{+0.1}}$ & ${13.1_{-0.6}^{+0.6}}$ \\
 53653.1 & BON & 12.0 & $144\pm7$ & ${-6.7_{-0.2}^{+0.2}}$ & ${11.6_{-0.7}^{+0.8}}$ \\
 53752.6 & BON & 6.3 & $153\pm6$ & ${-6.7_{-0.5}^{+0.4}}$ & ${9.9_{-1.1}^{+1.3}}$ \\
 53843.5 & BON & 2.3 & $146\pm7$ & ${-6.7_{-0.7}^{+0.6}}$ & ${13.3_{-1.7}^{+2.1}}$ \\
 53925.8 & BON & 5.5 & $159\pm7$ & ${-6.6_{-0.6}^{+0.5}}$ & ${13.8_{-1.9}^{+2.4}}$\\
 54023.9 & BON & 3.5 & $147\pm7$ & ${-6.8_{-0.7}^{+0.6}}$ & ${14.7_{-1.9}^{+2.4}}$\\
 54108.4 & BON & 6.6 & $149\pm6$ & ${-8.3_{-0.5}^{+0.5}}$ & ${12.3_{-1.4}^{+1.7}}$\\
 54190.0 & BON & 8.0 & $159\pm6$ & ${-9.2_{-0.7}^{+0.6}}$ & ${11.8_{-1.3}^{+1.7}}$\\
 54281.8 & BON & 6.5 & $153\pm8$ & ${-6.9_{-0.7}^{+0.6}}$ & ${9.4_{-1.1}^{+1.3}}$\\
 54614.3 & ASP & 0.8 & {---} & ${-10.9_{-0.6}^{+0.6}}$ & ${8.2_{-0.6}^{+0.6}}$\\
 54763.9 & ASP & 1.7 & {---} & ${-12.2_{-0.6}^{+0.6}}$ & ${7.4_{-0.4}^{+0.4}}$\\
 54863.6 & ASP & 1.5 & {---} & ${-12.2_{-0.8}^{+0.8}}$ & ${6.0_{-0.3}^{+0.3}}$\\
 54987.3 & ASP & 1.1 & {---} & ${-11.0_{-1.9}^{+1.7}}$ & ${6.6_{-0.5}^{+0.6}}$\\
 55046.1 & ASP & 1.2 & {---} & ${-12.7_{-1.4}^{+1.3}}$ & ${6.2_{-0.3}^{+0.3}}$\\
 56012.4 & PUPPI & 2.1 & $150\pm2$ & ${-19.0_{-0.8}^{+0.8}}$ & ${2.9_{-0.1}^{+0.1}}$ \\
 56087.2 & PUPPI & 2.1 & $155\pm3$ & ${-18.3_{-1.1}^{+1.0}}$ & ${2.7_{-0.1}^{+0.1}}$ \\
 56145.1 & PUPPI & 2.2 & $156\pm3$ & ${-21.0_{-1.2}^{+1.1}}$ & ${2.5_{-0.1}^{+0.1}}$ \\
 56178.0 & PUPPI & 1.8 & $150\pm3$ & ${-19.7_{-1.3}^{+1.1}}$ & ${2.3_{-0.1}^{+0.1}}$ \\
 56213.9 & PUPPI & 2.2 & $152\pm3$ & ${-19.4_{-1.1}^{+1.0}}$ & ${2.1_{-0.1}^{+0.1}}$ \\
 56263.7 & PUPPI & 1.9 & $149\pm3$ & ${-19.0_{-1.4}^{+1.3}}$ & ${1.9_{-0.1}^{+0.1}}$ \\
 56311.6 & PUPPI & 2.2 & $157\pm3$ & ${-18.2_{-1.1}^{+1.0}}$ & ${1.7_{-0.1}^{+0.1}}$ \\
 56347.5 & PUPPI & 2.2 & $153\pm3$ & ${-19.9_{-1.3}^{+1.2}}$ & ${1.8_{-0.1}^{+0.1}}$ \\
 56523.0 & PUPPI & 1.8 & $154\pm3$ & ${-19.3_{-2.6}^{+17.5}}$ & ${1.0_{-0.1}^{+7.4}}$ \\
 56583.9 & PUPPI & 2.2 & $162\pm3$ & ${-19.9_{-1.8}^{+1.5}}$ & ${0.6_{-0.1}^{+0.0}}$ \\
 56743.4 & PUPPI & 1.6 & $155\pm3$ & ${-19.5_{-3.3}^{+3.0}}$ & ${0.06_{-0.03}^{+0.04}}$ \\
 56758.4 & PUPPI & 1.5 & $148\pm3$ & ${-18.9_{-4.4}^{+4.3}}$ & ${0.04_{-0.04}^{+0.02}}$ \\
 56788.3 & PUPPI & 2.0 & $149\pm4$ & ${-13.4_{-4.6}^{+3.5}}$ & ${0.01_{-0.02}^{+0.02}}$ \\
 56819.2 & PUPPI & 1.9 & $150\pm4$ & ${-14.4_{-6.8}^{+18.7}}$ & ${0.00_{-0.03}^{+0.02}}$ \\
 56940.9 & PUPPI & 2.1 & $154\pm3$ & ${-15.6_{-7.7}^{+9.0}}$ & ${-0.34_{-0.05}^{+0.06}}$ \\
 56962.8 & PUPPI & 2.2 & $154\pm2$ & {---} & ${-0.44_{-0.04}^{+0.04}}$ \\
 57190.2 & PUPPI & 1.7 & $153\pm4$ & {---} & ${-1.3_{-0.1}^{+0.3}}$ \\
 57206.2 & PUPPI & 1.9 & $153\pm4$ & {---} & ${-1.3_{-0.2}^{+0.3}}$ \\
 57310.9 & PUPPI & 2.1 & $158\pm5$ & {---} & ${-2.0_{-0.1}^{+0.1}}$ \\
 57333.8 & PUPPI & 2.1 & $153\pm6$ & {---} & ${-2.0_{-0.1}^{+0.1}}$ \\
 57553.2 & PUPPI & 2.2 & $148\pm7$ & {---} & ${-3.2_{-0.1}^{+0.1}}$ \\
 57712.8 & PUPPI & 1.8 & $157\pm3$ & {---} & ${-3.7_{-0.2}^{+0.4}}$ \\
 57755.6 & PUPPI & 1.8 & $153\pm3$ & {---} & ${-3.7_{-0.4}^{+0.6}}$ \\
 57769.6 & PUPPI & 1.4 & $153\pm3$ & {---} & ${-4.0_{-0.2}^{+0.5}}$ \\
 57822.5 & PUPPI & 1.3 & $152\pm3$ & {---} & ${-4.4_{-0.2}^{+0.5}}$\\
 57858.4 & PUPPI & 1.8 & $152\pm5$ & {---} & ${-3.9_{-0.4}^{+0.7}}$ \\
 57890.3 & PUPPI & 1.3 & $149\pm6$ & {---} & ${-4.2_{-0.3}^{+0.5}}$ \\
 57921.2 & PUPPI & 2.1 & $152\pm3$ & {---} & ${-4.7_{-0.3}^{+0.7}}$ \\
 58116.7 & PUPPI & 1.2 & $154\pm3$ & {---} & ${-5.7_{-0.4}^{+0.8}}$ \\
 58174.5 & PUPPI & 1.5 & $153\pm3$ & {---} & ${-5.6_{-0.3}^{+0.7}}$ \\
 58207.5 & PUPPI & 2.0 & $153\pm3$ & {---} & ${-6.1_{-0.4}^{+0.7}}$ \\
 58237.4 & PUPPI & 2.0 & $151\pm2$ & {---} & ${-5.2_{-1.0}^{+0.7}}$\\
 58290.2 & PUPPI & 1.4 & $152\pm3$ & {---} & ${-6.2_{-0.4}^{+0.7}}$
\end{longtable}